\documentclass[aip,jmp]{revtex4-1}
\usepackage{amsmath}
\usepackage{amssymb}

\usepackage{graphicx}
\DeclareGraphicsExtensions{.pdf,.png,.jpg}

\begin{document}
\author{Bruno Mera}
\affiliation{CeFEMA, Instituto Superior
T\'ecnico, Universidade de Lisboa, Av. Rovisco Pais, 1049-001 Lisboa, Portugal}
\affiliation{Departamento de F\'{\i}sica, Instituto Superior
T\'ecnico, Universidade de Lisboa, Av. Rovisco Pais, 1049-001 Lisboa, Portugal}
\affiliation{Instituto de Telecomunica\c{c}\~oes, 1049-001 Lisbon, Portugal}
%
%
\title{Topological response of gapped fermions to a $\text{U}(1)$ Gauge field}
\begin{abstract}
We present a detailed path integral derivation of the topological response of gapped free fermions, in $2+1$ dimensions, to an external $\text{U}(1)$ Gauge field. The well-known Hall response is obtained by identifying the Chern-Simons term in the effective action with the correct coefficient. We extend the result to $2d+1$ dimensions in which the response is associated to a Chern-Simons term with a coefficient related to a characteristic class coming from topological band theory. We comment on the bulk-to-boundary principle which arises naturally when one considers the theory on a manifold with boundary.
\end{abstract}

\maketitle
\tableofcontents
\section*{Introduction}
\hspace{0.5cm} When a slab of a $2$-dimensional electron gas at low temperatures is subject to an external electric field, there exists a current flow in a direction perpendicular to the it in the plane, linear in the field, with the coefficient being quantized in units of $e^2/h$ -- the integer quantum Hall effect (IQHE). The quantization of the Hall conductivity in the IQHE is one of the hallmarks of topological phases of matter.  Using linear response theory, Thouless \textit{et al.} \cite{Thouless.1982} derived a formula for the transverse conductivity in terms of a topological invariant of the occupied bands of the system. The IQHE response can be modelled by an effective field theory for the associated Gauge field which is of the Chern-Simons type. This coupling is inherently topological in the sense that it does not depend on any choice of a Riemannian metric structure in the base space-time manifold. The Chern-Simons action functional describes an exactly solvable topological quantum field theory, with the correlation functions being topological invariants, namely, knot invariants of the $3$-manifold as discussed in Edward Witten's seminal paper\cite{Witten.1989}. The Chern-Simons action  appears, in addition to the usual Maxwell term, as an additional term, allowed by Gauge-invariance, for the bulk theory. In $3+1$ dimensions this term simply does not exist. The Chern-Simons coupling on a manifold with boundary is not Gauge invariant -- it has an ``anomaly''. The anomaly is cancelled by coupling the $\text{U}(1)$ Gauge field to chiral edge modes, rendering the theory consistent. This is an example of a general phenomenon occurring in topological phases of matter, where one finds an anomaly inflow from the bulk to the boundary. In fact, symmetry-protected topological phases (SPT) arising in free fermion systems can be understood from the standpoint of anomalies and anomaly inflow, as was recently shown by Edward Witten\cite{Edward.Witten.2015}. This was first conjectured by Furusaki \textit{et al.} in \cite{Furusaki.2013}, where several results towards this idea were obtained. The fact that topological phases are related to quantum anomalies in field theory incorporates the notion that topological phases of matter are insensitive to the microscopic details of the theory, including adding interactions consistent with the symmetries as long as the bulk gap is not closed.\\
\indent The Chern-Simons coupling appears  perturbatively when one considers $2+1$ dimensional QED and considers the one loop effective action (see for example \cite{Dunne.1999}). More generally, in $2d+1$ dimensions, general Chern-Simons terms appear when considering Wilson fermions on an hypercubic lattice \cite{Golterman.1992}, the coefficient being proportional to the winding number of a map from the Brillouin zone to the sphere associated with the fermion propagator. When the Wilson fermions have a mass coupling to a domain wall, the low energy effective theory can be described by chiral fermions bound to the $2d$ dimensional domain wall. The divergence of the Chern-Simons current exactly reproduces the anomaly of the chiral fermion zeromode bound to the domain wall. This is precisely the kind of phenomena occurring in topological phases of matter and described in the previous paragraph. The latter specific case is similar to what we will describe in the paper, but in the general case of gapped fermions.\\
 \indent In this paper we provide a detailed path integral derivation of the topological response of gapped fermions to an external $\text{U}(1)$ Gauge field in $2d+1$ dimensions. The effective field theory is of the Chern-Simons type with a coefficient  being a topological invariant. This topological invariant can be described by the single particle Green's function or, by integrating over the frequency, by the Berry curvature associated to the occupied Bloch bundle described in topological band theory. The formula obtained for the effective topological field theory describing the response of the system is consistent with that obtained by Xiao-Liang Qi \textit{et al.} in \cite{Qi.2008}. The method used in the present manuscript to approach the path integral uses a phase-space Wigner representation where position and momenta are treated in equal footing. This method was used by G. E. Volovik and V. M. Yakovenko in \cite{Volovik.1989}, when deriving the Chern-Simons term for the superfluid $^{3}$He-A thin film, where the coefficient of the coupling is also given by a momentum space topological invariant expressed in terms of the Green's function for the problem.\\
\indent The present paper provides a detailed pedagogical review of topological band theory, the Berry connection and the relation to the classification of complex vector bundles, which can be of general interest for both physicists and mathematicians working in the field of topological phases of matter. The main point of this paper is to provide a detailed path integral derivation of the bulk-to-boundary principle for gapped free fermions (with a $\text{U}(1)$ charge symmetry), which is fundamental to topological phases of matter, naturally associated, in this framework, to anomalies in quantum field theory.\\
\indent The paper is organized as follows: in section \ref{section: I} we describe the considerations regarding the free Hamiltonian used, review notions from topological band theory applied to a translation invariant representative; in section \ref{section: II} we review the concept of Berry connection on the Bloch bundle, provide the relevant formulae for the curvature in a coordinate independent way and relate the Berry connection to a universal connection on the tautological vector bundle over the Grassmannian; in  section \ref{section: III} we briefly discuss the Wigner representation that will be used to compute the path integrals later on; in section \ref{section: IV} we compute coefficient of the Chern-Simons term in $2+1$ dimensions; in section \ref{section: V}, the result of the previous section is generalized to $2d+1$ dimensions; in \ref{section: VI} we briefly comment on the bulk-to-boundary principle arising from the previous derivations; finally, we present the conclusions. Supplementary material is provided in the appendices: in Appendix A, one can find a self-contained discussion of characteristic classes and Chern-Simons forms in the context of the present manuscript; on Appendix B a brief discussion of anomalies, namely gauge anomalies, is presented with focus on the descent relations relevant for the bulk-to-boundary principle appearing in topological phases of matter.
\section{Gapped free fermions and topological band theory}
\label{section: I}
\hspace{0.5cm} We consider the problem of determining the topological part of the response of gapped free fermions minimally coupled to an external $\text{U}(1)$ Gauge field in $2+1$ dimensions and later to generalize the result to $D+1\equiv 2d+1$ dimensions ($D=2d$ spatial dimensions and $1$ time dimension). For the discussion in this section and the two following ones (i.e. Sections \ref{section: I}, \ref{section: II} and \ref{section: II}) we take arbitrary $d$. In the absence of the Gauge field, the system is described, in second quantization, by a quadratic Hamiltonian in fermion creation and annihilation operators regularized on a lattice,
\begin{align}
\mathcal{H}=\sum_{i,j}\psi_i^{*}h_{ij}\psi_j,
\end{align}
where the sum goes over all the associated degrees of freedom, including the lattice sites. There is a trivially conserved $\text{U}(1)$ charge associated to $\mathcal{H}$, namely the particle number $Q=\sum_{i}\psi^{*}_i\psi_i$, generating transformations $\psi_j\rightarrow e^{i\alpha}\psi_j$ and $\psi^{*}_j\rightarrow \psi^{*}_j e^{-i\alpha}$ with real $\alpha$. This charge will be minimally coupled to a $\text{U}(1)$ Gauge field.\\
\indent Since we are interested in the topological part of the response, we can consider, without loss of generality, that the free Hamiltonian is translation invariant, namely, it has the form,
\begin{align}
\mathcal{H}=\int_{T^{D}} \frac{d^{D}k}{(2\pi)^D} \ \psi^{*}(k)\cdot H(k)\cdot \psi(k).
\end{align}
In the above formula, the summation is over the first Brillouin zone which topologically is a torus $T^{D}$. The quantity $\psi(k):=[\psi_{1}(k),..., \psi_{n}(k)]^{t}$ ($\psi^{*}(k):=[\psi_{1}^{*}(k),...,\psi_{n}^{*}(k)]$) is an array of fermion annihilation (creation) operators with $k\in  T^D$.  The assignment $H:T^{D}\ni k\mapsto H(k) \in \text{Herm}(\mathbb{C}^n)$ (where $\text{Herm}(\mathbb{C}^n)$ is the set of Hermitian matrices in $\mathbb{C}^n$) is assumed to be smooth. The integer $n$ is the number of bands. We assume that the Fermi level is at zero, and the gap condition is simply $\det H(k)\neq 0$ for every $k\in T^{D}$. One can also choose a representative such that the Hamiltonian has a flat spectrum.
Indeed, the matrix $H(k)$ can be locally diagonalized by a matrix $S(k)\in\text{U}(n)$, i.e.,
\begin{align}
H(k)=S(k)\cdot\text{diag}(\varepsilon_1(k),...,\varepsilon_n(k))\cdot S^{*}(k),
\end{align}
where $\varepsilon_i(k)$, $1\leq i\leq n$, are the eigenvalues of $H(k)$, which are smooth functions. Then, by taking $\varepsilon_{i}(t,k)=(1-t)\varepsilon_i(k)+t \varepsilon_i(k)/|\varepsilon_i(k)|$, $t\in [0,1]$, due to the assumption of the gap, provides a smooth deformation between any (translation invariant) representative and flat spectrum representative:
\begin{align}
H_t(k)=S(k)\cdot\text{diag}(\varepsilon_1(t,k),...,\varepsilon_n(t,k))\cdot S^{*}(k),\ t\in[0,1],
\end{align}
with $H_0(k)\equiv H(k)$ and
\begin{align}
H_1(k)=S(k)\cdot\left[\begin{array}{cc}
-I_r & 0\\
0 & I_{n-r}
\end{array}\right]\cdot S^{*}(k),
\end{align}
where $r$ is the number of bands below the Fermi level.\\
\indent The unitary matrix $S$ is only defined locally and up to right multiplication by a matrix in $\text{U}(r)\times\text{U}(n-r)$ which preserves the diagonal matrix $\text{diag}(-I_r,I_{n-r})$. Locally it assumes the form
\begin{align}
S(k)=[v_1(k),...,v_{r}(k),v_{r+1}(k),...,v_{n}(k)],
\end{align}
where $\{v_i(k)\}_{1\leq i\leq r}$ is an orthonormal basis for the fibre at $k\in T^{D}$ of the ``occupied Bloch bundle'' (with total space $E=\{(k,v)\in T^{D}\times\mathbb{C}^n: H(k)\cdot v=-v\}$ and the obvious projection onto the first factor $T^{D}$) and similarly $\{v_i(k)\}_{r+1\leq i\leq n}$ is an orthonormal basis for the fibre at $k\in T^{D}$ of the ``unoccupied bundle'' (the orthogonal complement bundle). The collection $\{v_i(k)\}_{1\leq i\leq n}$ yields an orthonormal basis for $\mathbb{C}^n$. We can also write,
\begin{align}
H(k)&=-\sum_{i=1}^{r}v_i(k)v_i^{*}(k) +\sum_{i=r+1}^{n}v_i(k)v_i^{*}(k)\\
&=-P(k)+ (I_n-P(k))=I_n -2P(k),\nonumber
\end{align}
where $*$ denotes the Hermitian conjugate. In fact, the function $H:k\mapsto H(k) $ defines a map $\phi:T^{D}\rightarrow \text{Gr}_{k}(\mathbb{C}^n)=\text{U}(n)/\text{U}(r)\times\text{U}(n-r)$. Since we are studying Hamiltonians up to an equivalence that is given by smooth deformation without closing the gap, we are interested in the homotopy class of $\phi$. The homotopy class of $\phi$ does not depend on the translation invariant representative that we choose for $H$. The space $\text{Gr}_k(\mathbb{C}^n)$ is the set of $k$-planes over $\mathbb{C}^n$ and it can be identified with the space of $\text{rank}$ $k$ orthogonal projections in $\mathbb{C}^n$. Therefore, the map $\phi: k\mapsto P(k)$ defines, uniquely, a map to $\text{Gr}_k(\mathbb{C}^n)$. In the absence of any additional generic symmetry, such as time reversal or particle hole symmetry, the homotopy class of the map $\phi$, for $n$ large enough, completely classifies the system or equivalently completely determines the isomorphism class of the occupied bundle \footnote{The condition of having $n$ large enough is necessary since otherwise there can be maps which are not homotopic but lead to isomorphic vector bundles. One can show that for $n$ large enough the set of isomorphism classes of complex vector bundles over a compact manifold $M$, $\text{Vect}_{r}(M)$, is isomorphic to the set of homotopy classes of maps from the manifold to the Grassmannian of $k$ planes in $\mathbb{C}^n$, i.e. $\text{Vect}_{r}(M)\cong [M,\text{Gr}_{r}(\mathbb{C}^n)]$, see for e.g. \cite{Atiyah.1989}. This leads to the good notion of topological equivalence of Hamiltonians -- gapped Hamiltonians are considered to be topologically equivalent if, by adding enough flat bands (hence the large $n$), they can be continuously deformed into each other \cite{Kitaev.2009}. This is related to the notion of stable equivalence in $K$-theory.}. The occupied bundle is simply the pullback bundle $E=\phi^{*}E_0$, where $E_0=\{(V,v)\in \text{Gr}_k(\mathbb{C}^n)\times\mathbb{C}^n:v\in V \}$ is the tautological vector bundle over the Grassmannian of $k$-planes on $\mathbb{C}^n$. 
\section{The Berry connection on the Bloch bundle, the curvature endomorphism and relation to the universal connection}
\label{section: II}
\hspace{0.5cm} In this section we will review the concept of the Berry connection on the occupied bundle $E$ and understand it from the point of view of the classifying map $\phi: T^{D}\rightarrow \text{Gr}_k(\mathbb{C}^n)$. First, we endow the trivial bundle $T^{D}\times\mathbb{C}^n$ with the flat connection $d$ ($d$ acts trivially on the global trivialization induced by the canonical basis of $\mathbb{C}^n$) and the Hermitian structure induced from the standard  Hermitian inner product in $\mathbb{C}^n$. Then, the occupied sub-bundle $E\subset T^{D}\times\mathbb{C}^n$ has a connection in a canonical way from the orthogonal projection $\nabla=Pd$. This is the so-called Berry connection. In terms of a local frame field $S=[v_1,...,v_r]$,
\begin{align}
\nabla (S)=P\cdot d(S)=(S\cdot S^{*})\cdot dS=S\cdot(S^{*}\cdot dS),
\end{align}
where $S^{*}$ is the Hermitian conjugate of $S$. Hence, in terms of $S$, the connection coefficients read $S^{*}\cdot dS$. For simplicity of notation, we will now drop the matrix multiplication symbol which will be implicitly understood. For instance, we will write $S^{*}dS$ instead of $S^{*}\cdot dS$. Now let $s$ be a section of $E$, i.e., $s\in \Gamma (E)$. In terms of the canonical global frame on the trivial bundle, $I$ (represented by the $n\times n$ identity matrix), if $s=I a\equiv a\in\Gamma(E)\subset \Gamma(T^2\times\mathbb{C}^n)$ with $a=[a^1,...,a^{n}]$ the array of the coordinate functions of $s$ in the global frame $I$,
\begin{align}
\nabla(s)=\nabla(I a)\equiv Pda,
\end{align}
where we identify $P$ with the matrix in the global trivialization (i.e. $P\cdot I\equiv I\cdot P\equiv P$). Since $s\in\Gamma(E)$, we have $Pa=a$, and hence
\begin{align}
d(Pa)=dP a+Pda=da\Rightarrow Pda &=da-dP a\\
&=da-(dP P)a.\nonumber
\end{align}
Therefore, in terms of this global frame in the trivial bundle, the connection coefficients can be interpreted as $-dPP$.\\
The curvature of the occupied bundle is locally given by
\begin{align}
\nabla^2(S):=(\nabla\wedge \nabla)(S)=S\cdot(dS^*\wedge dS+S^{*}dS\wedge S^{*}dS)=S\cdot(dS^{*}(I_k-SS^{*})\wedge dS),
\end{align}
thus, locally, the curvature coefficients are given by $dS^{*}(I_k-S S^{*})\wedge dS\equiv dS^{*} P^{\perp}\wedge dS$. \\
\indent The curvature can also be seen as an endomorphism of the trivial bundle $T^{D}\times\mathbb{C}^n$. Take $s=a\in \Gamma(E)\subset \Gamma(T^{D}\times \mathbb{C}^n)$ and notice that
\begin{align}
\nabla^2(s)&=\nabla\wedge (I(da -dP a))=I(P dP\wedge da)=I(P dP\wedge(P da+dP a))\\
&=I(PdP\wedge dP P) a,\nonumber
\end{align}
where we noticed that $PdP P\equiv 0$ for any projector. Hence, the curvature endomorphism (extended to act on the whole trivial bundle by considering its action on the projection onto $E$) is given by
\begin{align}
\Omega=P dP \wedge dP P.
\end{align}
We can also write, because $P^2=P$,
\begin{align}
\Omega&=P dP \wedge dP P= (-dP P+ dP)\wedge dP P=dP\wedge dP\wedge P\\
&=dP\wedge(dP-P dP)=dP(I-P) \wedge dP \nonumber \\
&=dPP^{\perp}\wedge dP. \nonumber
\end{align}
In terms of a local coordinate basis $(x^{1},x^2,..,x^{D})$ on the base manifold,
\begin{align}
\Omega_{ij}:=\Omega(\frac{\partial}{\partial x^i},\frac{\partial}{\partial x^j})=\frac{\partial P}{\partial x^i}P^{\perp}\frac{\partial P}{\partial x^j}-\frac{\partial P}{\partial x^j}P^{\perp}\frac{\partial P}{\partial x^i}.
\end{align}
We will now show that the Berry connection is the pullback connection from a``universal'' connection on $E_0$ equipped with a natural Hermitian structure coming from equipping $\mathbb{C}^n$ with the standard Hermitian structure. A local orthonormal frame field on $E_0$ corresponds to a local section of the bundle $\text{V}_k(\mathbb{C}^n)\rightarrow \text{Gr}_k(\mathbb{C}^n)$, namely $s:U\subset\text{Gr}_k(\mathbb{C}^n)\rightarrow  \text{V}_k(\mathbb{C}^n)$, with $V_{k}(\mathbb{C}^n)=\{f\in \mathbb{C}^{n\times k}: f^{*}f=I_k\}$ being the set of orthonormal $k$-frames (also known as the Stiefel manifold) with the projection $\pi: f\mapsto ff^{*}\in\text{Gr}_k(\mathbb{C}^n)$. The tangent space $T_f V_k(\mathbb{C}^n)$ at an arbitrary frame $f$ has a vertical subspace which corresponds to the tangent space at $f$ to the fibre $\pi^{-1}(ff^{*})=\{f\cdot U:U\in\text{U}(k)\}$,
\begin{align}
V_{f}=\{v\in\mathbb{C}^{n\times k}: v=f\cdot X, \text{ for some } X\in\mathfrak{u}(k)\}.
\end{align}
The space $\mathbb{C}^{n\times k}\supset V_k(\mathbb{C}^n)$ has a natural Hermitian structure ($\langle u,v\rangle =\text{tr}(u^{*}v)$), and, therefore, we can introduce, from the real part of this Hermitian inner product, a Riemannian metric in $V_k(\mathbb{C}^n)$. We then have a unique way to identify the horizontal subspace at $f$, take  $H_f=V_f^{\perp}$, so that $d_f\pi:H_f\subset T_f V_k(\mathbb{C}^n)\overset{\cong}{\rightarrow} T_{ff^{*}}\text{Gr}_k(\mathbb{C}^n)$ is an isomorphism -- this defines a connection in the Ehresmann sense, the universal connection. Explicitly,
\begin{align}
H_{f}&=\{v\in\mathbb{C}^{n\times k}: \text{Re}\ \text{tr}(v^{*}fX)=0,\ \text{for all } X\in\mathfrak{u}(k)\}\\
&=\{v\in\mathbb{C}^{n\times k}: v^{*}f-f^{*}v=0\}.\nonumber
\end{align}
Parallel transport of a frame $f$ along a curve $\gamma:[0,1]\rightarrow \text{Gr}_k(\mathbb{C}^n)$ of $k$-planes of $\mathbb{C}^n$ is then defined to be a lift $\widetilde{\gamma}:[0,1]\rightarrow \text{Gr}_k(\mathbb{C}^n)$ such that,
\begin{align}
\frac{d\widetilde{\gamma}}{dt}(t)\in H_{\widetilde{\gamma}(t)}, \text{ for all } t\in[0,1].
\end{align}
Equivalently,
\begin{align}
\frac{d\widetilde{\gamma}^{*}}{dt}(t)\widetilde{\gamma}(t)-\widetilde{\gamma}^{*}(t)\frac{d\widetilde{\gamma}}{dt}(t)=0, \text{ for all } t\in[0,1].
\end{align}
If we have a local frame field $s:U\subset \text{Gr}_k(\mathbb{C}^n)\rightarrow V_k(\mathbb{C}^n)$ such that $s(\gamma(0))=s(ff^{*})=f$ and, for the sake of simplicity, assuming $\gamma([0,1])\subset U$, we can write,
\begin{align}
\widetilde{\gamma}(t)=s(\gamma(t))\cdot U(t), \text{ for all } t\in[0,1].
\end{align}
for some function $U:[0,1]\rightarrow \text{U}(k)$ with $U(0)=I_k$. Write $\gamma^{*}s(t)\equiv s$ for simplicity of notation. We can then write the horizontality condition as,
\begin{align}
\frac{dU^{*}}{dt}s^{*}sU + U^{*}\frac{ds^*}{dt}s U -\text{H.c.}=0, \text{ for all } t\in[0,1].
\end{align}
or equivalently, using $s^{*}s=I_k$ and the derivative $\frac{ds^{*}}{dt}s +s^{*}\frac{ds}{dt}=0$,
\begin{align}
U\frac{dU^{*}}{dt}+\frac{ds^*}{dt}s=0, \text{ for all } t\in[0,1].
\end{align}
The group element $U(1)\in\text{U}(k)$ provides parallel transport of elements of $E_0$ also -- hence a universal connection on $E_0$. If $f$ is a unitary frame at $ff^*=\gamma(0)$, then $s(\gamma(1))\cdot U(1)$ is a frame at $\gamma(1)$. Hence, if we write $v\in (E_{0})_{\gamma(0)}$ as $v=f\cdot a$ (with $a=[a^1,...,a^k]$ being the coordinates of $v$ with respect to the frame $f$), then the image under parallel transport, $\tau_{\gamma}(v)\in (E_{0})_{\gamma(1)}$, is
\begin{align}
\tau_{\gamma}(v)=s(\gamma(1))\cdot U(1)\cdot a.
\end{align}
The local connection form, is immediately read off, in terms of the local frame field $s$
\begin{align}
\omega(\frac{d\gamma}{dt})=-\frac{dU}{dt}U^{*}=U\frac{dU^{*}}{dt}=-\frac{ds^{*}}{dt}s=s^{*}\frac{ds}{dt},
\end{align}
hence,
\begin{align}
\omega=s^{*}ds.
\end{align}
So we have a connection $\nabla:\Gamma(E_0)\rightarrow \Omega^{1}(\text{Gr}_k(\mathbb{C}^n),E_0)$, which for local sections $t\in\Gamma(E_0|_{U})\subset \Gamma(U\times \mathbb{C}^n)$ , $U\subset \text{Gr}_k(\mathbb{C}^n)$, written is terms of a local frame field $s$, $t=s\cdot a$, acts as
\begin{align}
\nabla t= s\cdot (da +\omega a)=s\cdot (da+s^{*}ds a)=ss^{*}d(s\cdot a)=Pdt.
\end{align}
Now if we have a map $\phi:T^{D}\rightarrow \text{Gr}_k(\mathbb{C}^n)$, we can define, uniquely, a pullback connection $\phi^{*}\nabla: \Gamma(\phi^{*}E_0)\rightarrow \Omega^1(T^{D},\phi^{*}E_0)$, by
\begin{align}
(\phi^{*}\nabla)(\phi^{*}s)=\phi^{*}(\nabla s), \text{ for all } s\in \Gamma(E_0).
\end{align}
The connection form in terms of a local frame field $s$ for $E_0$, reads $\omega=s^{*}ds$. This, in turn, induces a connection form in $\phi^{*}E_0$, explicitly,
\begin{align}
\phi^{*}\omega=\phi^{*}(s^* ds).
\end{align}
This connection is precisely the Berry connection as one can immediately see. Take the smooth map $\phi:T^{D}\rightarrow \text{Gr}_k (\mathbb{C}^n)$ and an open cover of $\text{Gr}_k(\mathbb{C}^n)$, $\{U_i\}_{i\in I}$, with associated local trivializations of $V_k(\mathbb{C}^n)$, $\{s_i:U_i\rightarrow V_k(\mathbb{C}^n)\}_{i\in I}$. Then, $\{\phi^{-1}(U_i)\}_{i\in I}$ is an open cover of $T^{2d}$ and $\{\phi^{*}s_i\}$ is an atlas of orthonormal trivializations of $E=\phi^{*}E_0$ (or the associated orthonormal frame bundle). Clearly, writing $S_i=\phi^{*}s_i$, $i\in I$,
\begin{align}
\nabla (S_i)=PdS_i=S_i\cdot (S_i^{*}dS_i)=\phi^{*}(s_i\cdot (s_i^{*}ds_i)), \text{ for any } i\in I.
\end{align}
Hence, the Berry connection on the Bloch bundle is the pullback connection of the universal connection defined on $E_0$. Notice that although the isomorphism class of $E$ depends only on the homotopy class of $\phi$, the specific form of the Berry connection depends explicitly on $\phi$. Characteristic classes (see Appendix A), though, such as the Chern classes, depend only on the homotopy class of $\phi$, or, equivalently, the isomorphism class of $E=\phi^{*}E_0$, therefore do not depend on the connection chosen.
\section{Wigner Transform of the Green's function}
\label{section: III}
\hspace{0.5cm} Because we are interested in coupling our theory minimally to an external Gauge field we will spoil the translation invariance of the original Green's function $G_0(x_1,x_2)=G_0(x_1+a,x_2+a)$, with $a=(a^1,...,a^{D+1})$. In the presence of an external $\text{U}(1)$ Gauge field the kinetic momentum of the electrons gets shifted,
\begin{align}
p_\mu\rightarrow p_\mu -eA_\mu(x),
\end{align} 
where $e$ is the electric charge. It would then be useful to have a phase space representation for the Green's function in which $x^\mu$ and $p_\mu$ are present and we can formally write the minimal coupling prescription and perform a Taylor expansion in the vector potential $A_\mu(x)$:
\begin{align}
G_0^{-1}(p)\rightarrow G^{-1}_{A}(p,x)=G_0^{-1}(p-eA(x))\approx G_0^{-1}(p) -eA_\mu(x)\frac{\partial G_0^{-1}}{\partial p_\mu}(p).
\end{align}
This is achieved through the Wigner transform. One considers a two point function (or, more generally, a distribution) $A(x_1,x_2)$ and performs a Fourier transform on the difference of variables, i.e. the relative coordinate,
\begin{align}
x^\mu:=x_1^\mu-x_2^\mu.
\end{align}
The remaining variable is the center of mass variable,
\begin{align}
X^\mu:=(x_1^\mu+x_2^\mu)/2.
\end{align}
The center of mass variable is a macroscopic variable that should be the scale at which the Gauge field varies. The relative coordinate should be of the order of $\lambda_F\approx 1/p_F$, the Fermi length. With respect to this variable, there should be no variation of the Gauge field.\\
\indent Explicitly,
\begin{align}
A(p,X)=\int d^{D+1}x \ e^{-ip\cdot x} A(X+\frac{x}{2},X-\frac{x}{2}).
\end{align} 
If the original function $A(x_1,x_2)$ was translation invariant, then the Wigner transform coincides with the Fourier transform. The Wigner transform of the product (convolution in real space) of two operators is given by the Moyal product expansion,
\begin{align}
\int d^{D+1}x \ A(x_1,x)B(x,x_2)\rightarrow A(p,X)B(p,X)+\frac{i}{2}\{A,B\}_{\text{PB}}(p,X) +...,
\end{align}
where $\{.,.\}_{\text{PB}}$ denotes the Poisson bracket in phase space.\\
\indent To show the power of the Wigner representation, consider the Hamiltonian, in first quantized form, given by,
\begin{align}
H=p^{2}/2 +V(x),
\end{align}
where $V$ is a potential. The differential operator associated with the Green's function of the problem is then
\begin{align}
G^{-1}:=\mathcal{D}=i\frac{\partial}{\partial t}-\frac{p^2}{2}-V(x).
\end{align}
More precisely,
\begin{align}
G^{-1}(x_1,x_2)=\left(i\frac{\partial}{\partial t_1}-\frac{p_1^2}{2}-V(x_1)\right)\delta^{3}(x_1-x_2),
\end{align}
In the Wigner representation,
\begin{align}
G^{-1}(p,X)=p_0-(p_i)^2/2 - V(X)=p_0- H(p_i,X^i).
\end{align}
Hence, $G^{-1}$ became a phase space function in which the classical Hamiltonian appears explicitly. Now, when we perform the minimal coupling to a Gauge field,
\begin{align}
\mathcal{D}\rightarrow \mathcal{D}_{A}=i\frac{\partial}{\partial t}- eA_0(x) -\frac{1}{2}(p_i-eA_i(x))^2-V(x)
\end{align}
In the approximation $A(x_1)=A(X+x/2)\approx A(X)$ the vector potential factors out of the integral in the Wigner transform and we can write,
\begin{align}
G^{-1}_A(p,X)\approx(p_0-eA_0(X))-(p_i-eA_i(X))^2/2-V(X).
\end{align}
We can proceed even further by Taylor expanding with respect to $A$ and write,
\begin{align}
G^{-1}_A(p,X)\approx G^{-1}(p-eA(x),X)= G^{-1}(p,X)-eA_\mu(X)\frac{\partial}{\partial p_\mu}G^{-1}(p,X).
\end{align}
In our case, since we are considering fermions on a lattice, we will consider the Green's function $G_0(p,X)\equiv G_0(p)=(p_0 I_{n}- H(p_i))^{-1}$, where $p_i$ is only defined modulo $2\pi $ (to be more precise, only defined modulo translation by elements of the reciprocal lattice). In the presence of a magnetic field $p_\mu\rightarrow p_\mu -e A_\mu(x)$. Notice that this replacement has a catch since now the $D$-momentum variables are periodic. The justification for this was done by Peierls \cite{Peierls.1933} and the resulting representation for the Green's function is an analogue of the Wigner representation in which the product of two operators is also given by a Moyal product expansion. The details of the justification of this argument do not concern us here and we will take this as an assumption. In analogy with the Wigner representation, we perform the expansion,
\begin{align}
G_{A}^{-1}(p,x)\approx G_{0}^{-1}(p-eA(x))\approx G_{0}^{-1}(p)-e A_{\mu}(x)\frac{\partial}{\partial p_\mu} G_{0}^{-1}(p).
\end{align}
Notice how the Gauge field $A_\mu(x)$ couples minimally to the current density $J_\mu =\partial G_0^{-1}/\partial p_\mu$, whose charge is naturally identified with the $\text{U}(1)$ electric charge identified in the Introduction.
\section{Determining the topological response by path integral in $2+1$ dimensions}
\label{section: IV}
\hspace{0.5cm} In this section $D=2$, i.e, $2d+1=3$. We wish to determine the response of the system with respect to an external $\text{U}(1)$ Gauge field, on a flat space-time without boundary. We consider the free partition function of the system
\begin{align}
Z_0=\int [D\psi][D\psi^{*}]\exp(iS(\psi,\psi^{*})),
\end{align}
with 
\begin{align}
S=\int d^3x \ \psi^{*}(x)(i\partial/\partial t -H)\psi(x)=\int \frac{d^3p}{(2\pi)^3}\ \psi^{*}(p)(p_0 I_n -H(p_i))\psi(p).
\end{align}
Up to constant factors, $Z_0$ is formally given by the functional determinant
\begin{align}
\text{Det}(G_0^{-1}).
\end{align}
When we couple to a $\text{U}(1)$ external Gauge field $A$,
\begin{align}
Z_0\rightarrow Z(A)=\int [D\psi][D\psi^*]\exp(iS(\psi,\psi^*,A))\propto \text{Det}(G_{A}^{-1}).
\end{align}
We want to perform an expansion in the Gauge field and its derivatives. Notice that $Z(A)$ must be Gauge invariant, therefore, we are only interested in terms which satisfy this condition. In particular, in $2+1$ dimensions, in lowest order in the field and its derivatives there is a possible Gauge invariant term other than the usual Maxwell term, namely, the Chern-Simons term: $(1/4\pi)\int A\wedge dA$. To be consistent with Gauge invariance this term has to have an integer coefficent. We will show that this coefficient is precisely the first Chern number of the Bloch occupied bundle.\\
\hspace{0.5cm} Let us consider an expansion of $\text{Det}(G_A^{-1})$, given the approximation regarding the minimal coupling we have written in the previous section,
\begin{align}
\text{Det}(G_A^{-1})\approx \text{Det}(G_0^{-1} +\Sigma),
\end{align}
with $\Sigma(p,x)=-e A_\mu(x)\partial G_0^{-1}/\partial p_\mu(p)$. Formally, we can write,
\begin{align}
\text{Det}(G_0^{-1} +\Sigma)=\text{Det}(G_0^{-1})\cdot \text{Det}(I+G_0\Sigma)=Z_0\cdot\text{Det}(I+G_0\Sigma),
\end{align}
So, the effective action is given by expanding,
\begin{align}
\log \text{Det}(I+G_0\Sigma)=\text{Tr}\log(I+G_0\Sigma)\approx \text{Tr}(G_0\Sigma) -\frac{1}{2}\text{Tr} (G_0\Sigma G_0\Sigma) +...
\end{align}
Now the products in the functional traces are convolutions, so we must use the Moyal product expansion in the Wigner representation. We only care about the quadratic term and, in particular, we are interested in the one being a function of $A_\mu(x)$ and $\partial_{\mu}A_\nu (x)$. This term is of the form
\begin{align}
&-\frac{1}{2}\frac{i}{2}\times 2\times \int d^3 x \frac{d^3p}{(2\pi)^3}\text{tr}\left(G_0(p)\Sigma(p,x) \{G_0,\Sigma\}_{\text{PB}}(p,x)\right)=\\
&=-i\frac{e^{2}}{2}\times \int d^3 x \frac{d^3p}{(2\pi)^3}\text{tr}\left(G_0(p)A_\mu(x)\frac{\partial G_0^{-1}}{\partial p_\mu}(p)\{G_0,A_\nu\frac{\partial G_0^{-1}}{\partial p_\nu}\}_{\text{PB}}(p,x)\right) \nonumber \\
&=i\frac{e^{2}}{2}\int d^3x A_{\mu}(x)\frac{\partial A_\nu}{\partial x^\lambda}(x)\int \frac{d^3p}{(2\pi)^3}\text{tr}\left(G_0(p)\frac{\partial G_0^{-1}}{\partial p_\mu}(p)\frac{\partial G_0}{\partial p_\nu}(p)\frac{\partial G_0^{-1}}{\partial p_\lambda}(p)\right).\nonumber
\end{align}
Next, we will focus on the coefficients,
\begin{align}
c^{\mu\nu\lambda}:=\int \frac{d^3p}{(2\pi)^3}\text{tr}\left(G_0(p)\frac{\partial G_0^{-1}}{\partial p_\mu}(p)\frac{\partial G_0}{\partial p_\nu}(p)\frac{\partial G_0^{-1}}{\partial p_\lambda}(p)\right).
\end{align}
From $G_0 G_0^{-1} =I_n\Rightarrow d G_0 =-G_0 dG_0^{-1}G_0$, we can write,
\begin{align}
c^{\mu\nu\lambda}=-\int \frac{d^3p}{(2\pi)^3}\text{tr}\left(G_0(p)\frac{\partial G_0^{-1}}{\partial p_\mu}(p)G_0(p)\frac{\partial G_0^{-1}}{\partial p_\nu}(p)G_0(p)\frac{\partial G_0^{-1}}{\partial p_\lambda}(p)\right)
\end{align}
If we define a $1$-form with values in the Lie algebra $\mathfrak{gl}(\mathbb{C}^n)$,
\begin{align}
\omega=G_0 d G_0^{-1}:=\sum_{\mu}\omega^{\mu}(p) dp_{\mu},
\end{align}
then,
\begin{align}
c^{\mu\nu\lambda}=-\int \frac{d^3p}{(2\pi)^3}\text{tr}\left[\omega^\mu(p)\omega^\nu (p)\omega^{\lambda}(p)\right].
\end{align}
Clearly, $c^{\mu\nu\lambda}$ is invariant under cyclic permutation of $(\mu,\nu,\lambda)$. Furthermore, since we only care about Gauge invariant contributions to the action, we only care about the fully skew contribution $(1/3!)\sum_{\sigma\in S^3}\text{sgn}(\sigma)c^{\sigma(\mu\nu\lambda)}$. Hence, we only care about the integral
\begin{align}
-\frac{1}{3!}\frac{1}{(2\pi)^3}\int \text{tr}\left(\omega\wedge \omega\wedge \omega\right).
\end{align}
 To compute the integral, it is useful to perform a Wick rotation, so that we can perform the frequency integral in a standard way by complex contour integration. In this case $G_0$ gets replaced by
\begin{align}
\mathcal{G}_0(p)=\left(-ip_0 I_n + H(p_i)\right)^{-1}=\frac{1}{(-ip_0-1)}P(p_i) +\frac{1}{(-ip_0 +1)}P^{\perp}(p_i)
\end{align}
and 
\begin{align}
\mathcal{G}_0^{-1}(p)=-ip_0 I_n + H(p_i)=-ip_0 I_n +(P^{\perp}(p_i)-P(p_i))=-ip_0 I_n +I_{n} -2P(p_i).
\end{align}
We then have,
\begin{align}
\omega=\mathcal{G}_0 d \mathcal{G}_0^{-1}=\mathcal{G}_0 \left(-idp_0-\sum_i 2\frac{\partial P}{\partial p_i}dp_i\right)=\mathcal{G}_0\left(-idp_0 -2 dP\right).
\end{align}
Using the cyclic property of the trace (and being careful with the signs in the wedge products), we can write,
\begin{align}
\text{tr}\left(\omega\wedge \omega\wedge \omega\right)=(-i)12 dp_0\wedge\text{tr}\left(\mathcal{G}_0^2dP\wedge\mathcal{G}_0 dP \right)
\end{align}
Simplifying,
\begin{align}
&(-i)12\times\text{tr}[\left(\frac{1}{(-ip_0-1)^2}P(p_i) +\frac{1}{(-ip_0 +1)^2}P^{\perp}(p_i)\right)dP\\
&\wedge \left(\frac{1}{(-ip_0-1)}P(p_i) +\frac{1}{(-ip_0 +1)}P^{\perp}(p_i)\right)dP]. \nonumber
\end{align}
Using the identity $PdPP=0$ and $P^{\perp}dP P^{\perp}=-P^{\perp}dP^{\perp} P^{\perp}=0$, we get,
\begin{align}
(-i)12\times[&\frac{1}{(-ip_0 +1)^2(-ip_0 -1)}\text{tr}\left(PdPP^{\perp}\wedge dP\right)\\
& +\frac{1}{(-ip_0 +1)(-ip_0 -1)^2}\text{tr}\left(P^{\perp}dPP\wedge dP\right) ]. \nonumber
\end{align}
Now using the same identities, for the first term, 
\begin{align}
\text{tr}\left(PdPP^{\perp}\wedge dP\right)=\text{tr}\left((P+P^{\perp})dPP^{\perp}\wedge dP\right)=\text{tr}(\Omega).
\end{align}
For the second term, using the cyclic property of the trace (taking care of the signs in the wedge product),
\begin{align}
\text{tr}\left(P^{\perp}dPP\wedge dP\right) &=-\text{tr}\left(dPP^{\perp}\wedge dPP\right)\\
&=-\text{tr}\left(dPP^{\perp}dP(P+P^{\perp})\right) \nonumber \\
&=-\text{tr}(\Omega).\nonumber
\end{align}
Hence, we get,
\begin{align}
-\frac{1}{3!}\int \text{tr}\left(\omega\wedge \omega\wedge \omega\right)&=-2(-i)\int dp_0 f(p_0)\int_{T^2}\text{tr}(\Omega)\\
&=4\pi\int dp_0 f(p_0) \int_{T^2} \text{tr}\left(\frac{i\Omega}{2\pi}\right) \nonumber \\
&=  \left(c_1(E)\cdot [T^2]\right) \times \left(4\pi \int f(p_0) dp_0\right), \nonumber
\end{align}
where
\begin{align}
f(p_0)\equiv \frac{1}{(-ip_0 +1)^2(-ip_0 -1)} -\frac{1}{(-ip_0 +1)(-ip_0 -1)^2},
\end{align}
and $c_1(E)\cdot [T^2]$ denotes the pairing of the first Chern class of $E$, $c_1(E)=[\text{tr}(i\Omega/2\pi)]$, on the fundamental homology class $[T^2]$.
The integral over $p_0$ can be solved by considering a complex contour which is a half circle based on the real line. The integral on the arc vanishes (since $f$ goes sufficiently fast to zero when $|p|\rightarrow \infty$) and, hence, the integral over the complex contour reproduces the integral over the real line. Using the Residue theorem,
\begin{align}
\int f(p_0) dp_0=2\pi i \times \text{Res}_i(f)=-\pi.
\end{align}
Hence,
\begin{align}
-\frac{1}{3!}\int \text{tr}\left(\omega\wedge \omega\wedge \omega\right)= \left(c_1(E)\cdot [T^2]\right) \times (-4\pi^2).
\end{align}
Thus,
\begin{align}
c^{\mu\nu\lambda}|_{\text{Gauge Inv.}}&=-\frac{1}{3!}\frac{1}{(2\pi)^2}\varepsilon^{\mu\nu\lambda} \int \text{tr}(\omega\wedge\omega\wedge\omega)=\varepsilon^{\mu\nu\lambda}\frac{1}{(2\pi)^3}\left(c_1(E)\cdot [T^2]\right) \times (-4\pi^2)\\
&=-\varepsilon^{\mu\nu\lambda}\frac{1}{2\pi}\left(c_1(E)\cdot [T^2]\right) \nonumber
\end{align}
Hence, modulo non-universal terms,
\begin{align}
\text{Det}(I+G_0\Sigma)&\approx \exp\left(\frac{ie^{2}}{2}c^{\mu\nu\lambda} \int d^3x A_\mu(x)\frac{\partial A_{\nu}}{\partial x^{\lambda}}(x)\right)\\
&=\exp\left[\left(\frac{i e^2}{4\pi}\times \left(c_{1}(E)\cdot [T^2]\right)\right)\times\int A\wedge dA\right]\equiv\exp(iW(A)), \nonumber
\end{align}
Hence, the effective action for the Gauge field is
\begin{align}
W(A)=2\pi\sigma_{\text{Hall}}\times \text{CS}(A),
\end{align}
with $\sigma_{\text{Hall}}=(e^{2}/2\pi) \times \left(c_1(E)\cdot[T^2]\right)$ and $\text{CS}(A)=(1/4\pi)\int A\wedge dA$ being the Chern-Simons coupling in $2+1$ dimensions. The response is therefore a IQHE type response, namely,
\begin{align}
\delta \text{CS}=\int \left(\frac{1}{2\pi}\delta A\wedge F\right)=\int d^3 x \  \varepsilon^{\mu\nu\lambda}\frac{1}{2}\frac{1}{2\pi} \delta A_{\mu}(x) F_{\nu\lambda}(x).
\end{align}
Hence, we get a current density
\begin{align}
\langle j^{\mu}(x)\rangle\equiv \frac{\delta W}{\delta A_{\mu}(x)}=\frac{1}{2}\sigma_{\text{Hall}} \varepsilon^{\mu\nu\lambda} F_{\nu\lambda}(x).
\end{align}
In components, the previous equation yields the charge density 
\begin{align}
&\langle j^{0}(x)\rangle \equiv \langle \rho (x)\rangle =\frac{1}{2}\sigma_{\text{Hall}}\varepsilon^{0ij}F_{ij}(x)=\sigma_{\text{Hall}}F_{12}(x)=\sigma_{\text{Hall}} B(x),
\end{align}
and the vector current density
\begin{align}
&\langle j^{i}(x)\rangle =\frac{1}{2}\sigma_{\text{Hall}}\left(\varepsilon^{i0j}F_{0j}(x)+\varepsilon^{ij0}F_{j0}(x)\right)=\sigma_{\text{Hall}}\varepsilon^{ij} E_j(x), \ i\in \{1,2\},
\end{align}
where $B(x)$ and $(E_{1}(x),E_{2}(x))$ are the magnetic and electric fields, respectively. A few comments are in order. We have implicitly assumed space-time was $\mathbb{R}^{1,2}$ and that the external $\text{U}(1)$ Gauge field came from a trivial bundle over this space-time. The derivation for the general case of arbitrary space-time manifold $M$ has to be done with more care. In the next section we generalize this result for arbitrary $d$, i.e., in $2d+1$ dimensions.
\section{Chern-Simons response in $2d+1$ dimensions}
\label{section: V}
\hspace{0.5cm} Here we take $d$ arbitrary, i.e. we look for the generalization in $D+1=2d+1$ dimensions. For an Abelian Gauge field $A$, the Chern-Simons form $Q_{2d+1}$ is given by (see Appendix A for details)
\begin{align}
Q_{2d+1}(A,F)=\frac{i^{d+1}}{(d+1)!(2\pi)^{d+1}}A\wedge F^{d}.
\end{align}
Notice that the physical Gauge field and curvature are really $-i e A$ and $-i e F$, respectively, but we will take care of the constants in the end. Effectively, we wish to find a term in the action which contains $A_{\mu_0}(x)\partial_{\mu_1}A_{\mu_2}(x)...\partial_{\mu_{2d-1}}A_{\mu_{2d}}(x)$. This means that we  are interested in the contribution
\begin{align}
\frac{1}{d+1}(-1)^{d}\text{Tr}\left((G_0\Sigma)^{d+1}\right)
\end{align}
And, regarding the Moyal expansion of the above expression, we collect the terms,
\begin{align}
\frac{2^{d}}{d+1}\times \left(\frac{i}{2}\right)^{d}(-1)^{d}\int \frac{d^{2d+1}p}{(2\pi)^{2d+1}}d^{2d+1}x \ \text{tr}\left[(G_0(p)\Sigma(p,x))\left(\{G_0,\Sigma\}_{\text{PB}}(p,x)\right)^{d}\right].
\end{align}
The factor of $2^{d}$ is combinatorial (it has to do with all possible permutations of the Poisson bracket inside the trace). We obtain,
\begin{align}
c^{\mu_0...\mu_d}\int d^{2d+1}x \ A_{\mu_0}(x)\partial_{\mu_1}A_{\mu_2}(x)...\partial_{\mu_{2d-1}}A_{\mu_{2d}}(x),
\end{align}
with 
\begin{align}
&c^{\mu_0...\mu_d}=\frac{2^{d}}{d+1}\times e^{d+1}\left(\frac{i}{2}\right)^{d}(-1)^{d+1}\\
&\times \int \frac{d^{2d+1}p}{(2\pi)^{2d+1}}\text{tr}\left(G_0(p)\frac{\partial G_0^{-1}}{\partial p_{\mu_0}}(p)\frac{\partial G_0}{\partial p_{\mu_1}}(p)\frac{\partial G_0^{-1}}{\partial p_{\mu_2}}(p)...\frac{\partial G_0}{\partial p_{\mu_{2d-1}}}(p)\frac{\partial G_0^{-1}}{\partial p_{\mu_{2d}}}(p)\right) \nonumber\\
&=-\frac{2^{d}}{d+1}\times e^{d+1}\left(\frac{i}{2}\right)^{d}\int \frac{d^{2d+1}p}{(2\pi)^{2d+1}}\text{tr}\left(G_0(p)\frac{\partial G_0^{-1}}{\partial p_{\mu_0}}(p)G_0(p)\frac{\partial G_0^{-1}}{\partial p_{\mu_1}}(p)...G_0(p)\frac{\partial G_0^{-1}}{\partial p_{\mu_{2d}}}(p)\right), \nonumber
\end{align}
where we have used, iteratively, $\{G_0,\Sigma\}_{\text{PB}}=e \partial_{\mu}A_{\nu}(\partial G_0/\partial p_\mu) (\partial G_0/\partial p_\nu)$ and $dG_0=-G_0 dG_0^{-1} G_0$. Since we are interested in the Gauge invariant contribution, we consider only the contribution $(1/(2d+1)!)\sum_{\sigma\in S_{2d+1}} c^{\sigma(\mu_0...\mu_{2d})}$, obtaining,
\begin{align}
\frac{2^{d}}{d+1}\times \frac{1}{(2d+1)!(2\pi)^{2d+1}}e^{d+1}\left(\frac{i}{2}\right)^{d}(-1)\int \text{tr}(\omega^{\wedge (2d+1)}),
\end{align}
with $\omega=G_0dG_0^{-1}$. Again, we perform a Wick rotation, in which $G_0^{-1}(p)\rightarrow \mathcal{G}_0^{-1}(p)=-ip_0 I_n +H(p_i)$. Recall that,
\begin{align}
\mathcal{G}_0d\mathcal{G}_0^{-1}=\mathcal{G}_0(-idp_0 I_n -2dP).
\end{align}
We are then interested in computing the $p_0$ integral of
\begin{align}
\text{tr}\left[\mathcal{G}_0(-idp_0 I_n -2dP)...\mathcal{G}_0(-idp_0 I_n -2dP)\right].
\end{align}
Using the cyclic property of the trace, this reduces to,
\begin{align}
(2)^{2d}(2d+1)(-idp_0)\wedge \text{tr}\left[\mathcal{G}_0(\mathcal{G}_0 dP)^{\wedge (2d)}\right]
\end{align}
Because of the identities $PdPP=0$ and $P^{\perp}dP P^{\perp}=0$, only two contributions arise in the trace,
\begin{align}
\text{tr}\left[\mathcal{G}_0(\mathcal{G}_0 dP)^{\wedge (2d)}\right]&=\frac{1}{(-ip_0-1)^{d+1}(-ip_0+1)^{d}}\text{tr}\left(P^2dP P^{\perp}\wedge dP P\wedge...\wedge P^{\perp}dPP\right)\\
&+\frac{1}{(-ip_0-1)^{d}(-ip_0+1)^{d+1}}\text{tr}\left(P^{\perp 2}dPP\wedge dPP^{\perp}\wedge...\wedge PdPP^{\perp}\right) \nonumber \\
&=\frac{1}{(-ip_0-1)^{d+1}(-ip_0+1)^{d}}\text{tr}\left(PdP P^{\perp}\wedge dP P\wedge...\wedge P^{\perp}dPP\right) \nonumber \\
&+\frac{1}{(-ip_0-1)^{d}(-ip_0+1)^{d+1}}\text{tr}\left(P^{\perp}dPP\wedge dPP^{\perp}\wedge...\wedge PdPP^{\perp}\right)\nonumber
\end{align}
Now, the first trace,
\begin{align}
\text{tr}\left(PdP P^{\perp}\wedge dP P\wedge...\wedge P^{\perp}dPP\right)=\text{tr}\left(\Omega^{\wedge d}\right),
\end{align}
while the second,
\begin{align}
\text{tr}\left(P^{\perp}dPP\wedge dPP^{\perp}\wedge...\wedge PdPP^{\perp}\right)&=-\text{tr}\left(dP^{\perp}P\wedge dP P^{\perp}\wedge...\wedge PdPP^{\perp}\right)\\
&=\text{tr}\left(dPP\wedge dP P^{\perp}\wedge...\wedge PdPP^{\perp}\right) \nonumber \\
&=-\text{tr}\left(P dP P^{\perp}\wedge dP ...\wedge PdPP^{\perp}\wedge dPP\right) \nonumber\\
&=-\text{tr}\left(\Omega^{\wedge d}\right) \nonumber
\end{align}
Hence we get,
\begin{align}
\frac{2^{d}}{d+1}\times e^{d+1}i^{d}\frac{(-1)(2)^{d}}{(2d)! (2\pi)^{2d+1}}\text{tr}\left(\Omega^{\wedge d}\right)\times I(d),
\end{align}
where 
\begin{align}
I(d)&\equiv \int_{-\infty}^{\infty} (-idp_0)\left(\frac{1}{(-ip_0 -1)^{d+1}(-ip_0+1)^{d}}-\frac{1}{(-ip_0 +1)^{d}(-ip_0+1)^{d+1}}\right).
\end{align}
To determine $I(d)$, we use the Residue theorem, which is valid since the denominators decay sufficiently fast for $d>1$. First, we re-write
\begin{align}
I(d)&=(-1)^d\int_{-\infty}^{\infty}dp_0\left(\frac{1}{(p_0-i)^{d+1}(p_0+i)^{d}}-\frac{1}{(p_0 -i)^{d}(p_0+i)^{d+1}}\right)\\
&=(-1)^d (2\pi i)\left[\frac{1}{d!}\frac{d^d}{d p_0 ^d}\left(\frac{1}{(p_0+i)^{d}}\right)\bigg|_{p_0=i}-\frac{1}{(d-1)!}\frac{d^{d-1}}{d p_0^{d-1}}\left(\frac{1}{(p_0+i)^{d+1}}\right)\bigg|_{p_0=i}\right] \nonumber \\
&=(-1)^d (2\pi i)\left[\frac{(-1)^{d}d(d+1)...(2d-1)}{d!}\frac{1}{(2i)^{2d}}-\frac{(-1)^{d-1}(d+1)...(2d-1)}{(d-1)!}\frac{1}{(2i)^{2d}}\right] \nonumber \\
&=(-1)^{d}\frac{(2\pi i)(2d-1)!}{(d-1)!d!}\frac{1}{2^{2d-1}}, \nonumber
\end{align}
hence,
\begin{align}
& \frac{2^{d}}{d+1}\times e^{d+1}i^{d}\frac{(-1)(2)^{d}}{(2d)! (2\pi)^{2d+1}}\text{tr}\left(\Omega^{\wedge d}\right)\times (-1)^{d}\frac{(2\pi i)(2d-1)!}{(d-1)!d!}\frac{1}{2^{2d-1}}=\\
&=\frac{1}{d+1}\times \left(\frac{(-1)^{d+1}ie^{d+1}}{(2\pi)^d(d)!}\right)\times \left[\frac{1}{d!}\text{tr}\left(\frac{i\Omega}{2\pi}\right)^{\wedge d}\right] \nonumber
\end{align}
which for $d=1$ reduces to
\begin{align}
\frac{i e^2}{4\pi}\times \text{tr}\left(\frac{i\Omega}{2\pi}\right),
\end{align}
as evaluated previously. Evaluating the momentum integral, yields,
\begin{align}
\frac{1}{d+1}\times \left(\frac{(-1)^{d+1}ie^{d+1}}{(2\pi)^d(d)!}\right)\times \left(\text{ch}(E)\cdot[T^{2d}]\right),
\end{align}
where $\text{ch}(E)=[\text{tr}\left( e^{i\Omega/2\pi}\right)]$ is the Chern character of the occupied bundle $E\rightarrow T^{2d}$. The result is that we get a response described by the effective action $W(A)$, with
\begin{align}
\exp(iW(A))=\exp\left[i\left((-1)^{d+1}\frac{e^{d+1}}{d+1}\right)\times \left(\text{ch}(E)\cdot[T^d]\right)\times\int A\wedge \frac{1}{d!}\frac{F^{d}}{(2\pi)^d}\right].
\end{align}
Notice that the Chern-Simons form was given explicitly by (notice that the connection $1$-form is $-ieA$)
\begin{align}
Q_{2d+1}(-ieA,-ieF)=\frac{i^{d+1}e^{d+1}}{(d+1)!(2\pi)^{d+1}}(-i)^{d+1}A\wedge F\wedge...\wedge F=\frac{e^{d+1}}{(d+1)d!(2\pi)^{d+1}}A\wedge F^{d}
\end{align}
Thus,
\begin{align}
\exp(iW(A))&=\exp\left[ 2\pi i(-1)^{d+1}\times\left(\text{ch}(E)\cdot[T^{2d}]\right)\times  \int Q_{2d+1}(-ieA,-ieF)\right] \\
&=\exp\left[2\pi i\left(\text{ch}(E)\cdot[T^{2d}]\right)\times  \int Q_{2d+1}(ieA,ieF)\right]. \nonumber
\end{align}
We write the general result obtained in a compact way as,
\begin{align}
W(A)=2\pi \left(\text{ch}(E)\cdot[T^{2d}]\right)\times  \int_{M^{2d+1}} Q_{2d+1}(ieA,ieF).
\end{align}
\section{Bulk-to-boundary principle}
\label{section: VI}
\hspace{0.5cm} The reasoning behind detecting topological phases through anomalies goes as follows. We have seen that the effective action for the response of gapped free fermions to an external $U(1)$ gauge field $A$ is of the Chern-Simons type $Q_{2d+1}(A,F)$, namely $W(A)= 2\pi k \int _{X} Q_{2d+1}(A,F)$ where $X$ is the $2d+1$ space-time manifold and $k=\text{ch}(E)\cdot[T^{2d}]$. The gauge variation of this action is, by the descent relation (see Appendix B),
\begin{align}
s Q_{2d+1}(A,F)\equiv s Q_{2d+1}^{0}(A,F)=-dQ^{1}_{2m}(A,F),
\end{align}
meaning that if $X$ is a manifold with boundary $\partial X=Y$,
\begin{align}
s W(A)= 2\pi k \int_{X} s Q_{2d+1}(A,F)=-2\pi k \int_{X} dQ^{1}_{2m}(A,F)=-2\pi k\int_{\partial X=Y} Q^{1}_{2m}(A,F),
\end{align}
where we have used Stokes' theorem in the last step. This anomalous variation can only be cancelled by degrees of freedom living on the boundary $Y$ whose coupling to the gauge field $A$ also has an anomalous variation matching the previous. In $2d+1=3$ dimensions, the topological invariant $k$ reduces to the integer $c_1(E)\cdot[T^2]$, and the anomaly in the variation of the effective action is consistent with $|k|$-Weyl fermions living in the boundary coupled to $A$ (the handedness depending on the sign of $k$). These degrees of freedom are gapless and express the bulk-to-boundary principle of topological phases of matter.
\section{Conclusions}
\hspace{0.5cm} We have presented a detailed path integral derivation of the topological response of gapped free fermions to an external $\text{U}(1)$ gauge field in $2+1$ dimensions and later generalized to $2d+1$ dimensions.  The method relies on choosing a translation invariant representative for the Hamiltonian regularized on a lattice. This representative, in turn, allows us to use the results of topological band theory expressing invariants of the gapped topological phase in terms of the Berry curvature which appear later when performing the path integral using a Wigner representation in which momentum and position are treated on an equal footing.  The price to pay for the latter, is that convolutions, which on momentum space are simply products, are replaced by Moyal product expansions -- an expansion on Poisson brackets. The latter is exactly what allows us to obtain the relevant derivatives on the Gauge field to arrive at the Chern-Simons couplings allowed, in $2d+1$ dimensions, by Gauge invariance with a coefficient written in terms of the Berry curvature. The fact that the coefficient of the Chern-Simons coupling, a topological invariant, can also be written in terms of the single particle Green's function for the problem, roughly $\int_{\mathbb{R}\times T^{2d}} \text{tr}(\omega^{\wedge(2d+1)})$ with $\omega=\mathcal{G}d\mathcal{G}^{-1}$, depending only on the homotopy class of $\mathcal{G}^{-1}$ or, as denoted in Appendix A, $K(H):\mathbb{R}\times T^{2d}\rightarrow \text{GL}(\mathbb{C}^n)$, also confirms the stability with respect to interactions [We remark that in Appendix $A$ we write the expression in more geometrical terms by writing the integrand as a pullback by $K(H)$ of a winding class of the group $\text{GL}(\mathbb{C}^n)$]. In fact, this expression in terms of the Green's functions allows for the determination of this topological invariant whenever an effective free fermion Hamiltonian (even if not static) description is available coming, for example, from a self-energy correction to the inverse of the Green's function. We commented on the bulk-to-boundary principle which follows naturally as a consequence of the result obtained. On reviewing concepts of topological band theory we have also remarked the relation of the Berry connection and the universal connection built on the tautological vector bundle over the Grassmannian manifold.
\acknowledgements{B. M. acknowledges the support from DP-PMI and FCT (Portugal) through the grant SFRH/BD/52244/2013. B. M. acknowledges very fruitful discussions with Jo\~{a}o P. Nunes regarding the mathematical background of the manuscript.}
\section*{Appendix A: Characteristic Classes and Chern-Simons Forms}
\label{Appendix A}
\hspace{0.5cm} For the purpose of this paper, characteristic classes are de Rham co-cycles (i.e. closed differential forms modulo exact forms) describing the twisting of an isomorphism class of a (complex) vector bundle $\pi:E\rightarrow M$. If one introduces a connection $\nabla$ on $E$, then these de Rham co-cyles are represented by invariant polynomials on the curvature of the connection, hence they are even degree in cohomology. It is a classical result that these classes do not depend on the choice of connection. The discussion presented here follows Luis Alvarez Gaum\'{e}\cite{Alvarez.Gaume.1985} and Shigeyuki Morita\cite{Morita.2001} in parallel. Let $\omega=[\omega_{ij}]_{1\leq i,j \leq k=\dim E}$ and $\Omega=[\Omega_{ij}]_{1\leq i,j\leq k}$ be, respectively, a matrix of $1$-forms and $2$-forms representing locally, with respect to some local frame field on $E$ defined over an open set $U\subset M$, the connection and the associated curvature. Recall that $\Omega=d\omega +\omega\wedge \omega$. For the discussion of characteristic classes, it suffices to consider the polynomials
\begin{align}
P_m(\Omega)=\frac{1}{m!}\left(\frac{i}{2\pi}\right)^{m}\text{tr}\ \Omega^{m},
\end{align}
since all other invariant polynomials can be obtained as sums and wedge products of these. It is clear that the above formula, by conjugation invariance, defines an even degree differential form over $M$. The closure of $P_m$ follows from the cyclic property of the trace and the Bianchi identity $d\Omega+[\omega,\Omega]=d\Omega +\omega\wedge\Omega -\Omega\wedge\omega =0$,
\begin{align}
dP_m(\Omega)&=\frac{1}{(m-1)!}\left(\frac{i}{2\pi}\right)^{m}\text{tr}\left(d\Omega\wedge \Omega^{m-1}\right)\\
&=\frac{1}{(m-1)!}\left(\frac{i}{2\pi}\right)^{m}\text{tr}\left(\omega\wedge \Omega^{m}-\Omega\wedge\omega\wedge\Omega^{m-1}\right)=0 \nonumber
\end{align}
To show that these polynomials to not depend on the connection, it is standard to introduce an homotopy operator. The construction goes as follows. There is a natural map $\pi\times \text{id}: E\times \mathbb{R}\rightarrow M\times \mathbb{R}$ yielding a vector bundle over $M\times\mathbb{R}$ with the $\mathbb{R}$ factor in the base space parameterized by a coordinate $t$. The fibre of $E\times \mathbb{R}$ over an arbitrary point $(p,t)\in M\times \mathbb{R}$ is $(\pi\times\text{id})^{-1}\{(p,t)\}=\{(v,t)\in E\times \mathbb{R}: \pi(v)=p\}=E_{p}\times \{t\}\cong E_p$, i.e. a copy of $E_p$: the fibre over $p\in M$ of $E$. A section of $E$ can be seen as a section of $E\times \mathbb{R}$ (i.e. $\Gamma(E)\subset \Gamma(E\times \mathbb{R})$), by associating $\Gamma(E)\ni s\leftrightarrow s\times\text{id}\in \Gamma(E\times \mathbb{R})$ (notice that $s\times\text{id}: M\times\mathbb{R} \ni(p,t)\mapsto (s(p),t)\in E\times\mathbb{R}$). There is a global vector field $\partial/\partial t \in\Gamma(T(M\times\mathbb{R}))$ trivializing the $\mathbb{R}$ component of the fibre of the tangent bundle of $M\times\mathbb{R}$. Finally, a vector field $X\in\Gamma(TM)$ can be seen as a vector field in $\Gamma(T(M\times \mathbb{R}))=\Gamma(\text{pr}_1^{*}TM\oplus \text{pr}_{2}^{*} T\mathbb{R})$ by $\Gamma(TM)\ni X\leftrightarrow (\text{pr}_1^{*}X,0)\in \Gamma(T(M\times\mathbb{R}))$, where $\text{pr}_i$, $i=1,2$, are the natural projections onto $M$ and $\mathbb{R}$, respectively . Then, if we have two connections $\nabla^0$ and $\nabla^1$ on $E$, we can introduce an associated connection $\overline{\nabla}$ on $E\times \mathbb{R}$ with the following prescription:
\begin{itemize}
\item[(i)] $\overline{\nabla}_{\frac{\partial}{\partial t}} s\equiv 0$ if $s\in \Gamma(E)\subset \Gamma(E\times \mathbb{R})$;
\item[(ii)]$(\overline{\nabla}_{X} s) (p,t)=(1-t)(\nabla_{X}^{0}s)(p,t)+t(\nabla_{X}^{1}s)(p,t)$ for $s\in \Gamma(E)\subset \Gamma(E\times \mathbb{R})$ and $X\in \Gamma(TM)\subset \Gamma(T(M\times\mathbb{R}))$, and any $(p,t)\in M\times\mathbb{R}$ (notice that, on the RHS, we have used again $\Gamma(E)\subset \Gamma(E\times\mathbb{R})$ namely by taking $\nabla^{i}_X s\in\Gamma(E)\subset \Gamma(E\times\mathbb{R})$, $i=0,1$).
\end{itemize}
It is easy to check that the above conditions completely determine the connection $\overline{\nabla}$. In fact, if $s=[s_1,...,s_k]$ is a frame field for $E|_{U}$, where $U$ is an open set on $M$, then, $\overline{s}=[s_1,...,s_k]$ is a frame field for $(E\times\mathbb{R})|_{U\times\mathbb{R}}$, with the $s_i$'s now seen as sections of $E\times \mathbb{R}$, and,
\begin{align*}
\overline{\nabla}\overline{s}=\overline{s}\cdot \overline{\omega}, \text{ with } \overline{\omega}=(1-t)\text{pr}_1^{*}\omega^{0}+t\text{pr}_1^{*}\omega^{1},
\end{align*}
in which $\omega^{0}$ and $\omega^{1}$ are the local connection forms with respect to $s$ for the connections $\nabla^{0}$ and $\nabla^{1}$, respectively, i.e., $\nabla^{i}s=s\cdot \omega^{i}$, $i=0,1$. The local form of the curvature is readily determined by the equation $\overline{\Omega}=d\overline{\omega}+\overline{\omega}\wedge\overline{\omega}$.\\
\indent In fact, the continuous interpolation between two connections,
\begin{align}
\nabla_t=(1-t)\nabla_0 +t \nabla_1,\ t\in\mathbb{R},
\end{align}
is, as one can easily prove, a connection on $E$ for each $t\in \mathbb{R}$. The differential forms $\omega_t=(1-t)\omega_0 +t\omega_1$ and $\Omega_t$, $t\in \mathbb{R}$, are the local forms of the associated connection forms and curvature forms, respectively. The natural inclusion map $i_{t}: M\ni p\mapsto (p,t)\in M\times\mathbb{R}$, of course, gives
\begin{align}
i_t^{*}\overline{\omega}=\omega_t \text{ and } i_t^{*}\overline{\Omega}=\Omega_t,\ t\in\mathbb{R}.
\end{align}
Hence, we understand, 
\begin{align}
i_{t}^{*}P_m(\overline\Omega)=P_m(\Omega_t), \ t\in \mathbb{R}.
\end{align}
Since $i_0$ and $i_1$ are clearly homotopic, by homotopy invariance of cohomology, it is now clear that $P_m(\Omega_0)$ and $P_m(\Omega_1)$ represent the same de Rham cocycle. It is useful to see their specific relation as differential forms by means of the homotopy operator
\begin{align*}
\Phi:\Omega^{\bullet}(M\times \mathbb{R}) \ni\eta\mapsto \int_{0}^{1}dt \ \iota(\frac{\partial}{\partial t})\eta\in \Omega^{\bullet -1}(M),
\end{align*} 
which one can easily show to satisfy,
\begin{align}
i_{1}^{*}\eta -i_{0}^{*}\eta= (d\Phi +\Phi d)\eta, \text{ for any } \eta\in\Omega^{\bullet}(M\times\mathbb{R}).
\end{align}
Notice that after performing the interior product $\iota(\partial/\partial t)$, one identifies the differential form appearing under the sign of integral with the pullback by $i_t:M\hookrightarrow M\times \mathbb{R}$ to have a differential form on $M$.\\
\indent Taking $\eta\equiv S(\overline{\omega},\overline{\Omega})$, an arbitary polynomial of the connection and curvature, we get,
\begin{align}
S(\omega_1,\Omega_1) -S(\omega_0,\Omega_0)= (d\Phi +\Phi d)S(\overline{\omega},\overline{\Omega}).
\end{align}
In particular, taking,
 $\eta= S(\overline{\omega},\overline{\Omega})\equiv P_m(\Omega_t)$, we get,
\begin{align}
&P_m(\Omega_1)-P_m (\Omega_0)=d\Phi(P_m(\overline{\Omega})),\\
& \text{ with } \Phi(P_m(\overline{\Omega})) =\frac{1}{(m-1)!}\left(\frac{i}{2\pi}\right)^{m}\int_{0}^{1}dt \ \text{tr} \left((\omega_1-\omega_0)\wedge \Omega_t^{m-1}\right)\nonumber
\end{align}
hence, the de Rham co-cycle associated to $P_m(\Omega)$ does not depend on the connection, as claimed previously. On a single patch $U$, one can consider the trivial connection (with respect to the local trivialization considered) and a reference connection $\nabla$ with local connection form $\omega$. Then, the path of connections over this patch is $\omega_t= t\omega$. The previous homotopy formula gives,
\begin{align}
P_m(\Omega)=d Q_{2m-1}(\omega, \Omega),
\end{align} 
with
\begin{align}
Q_{2m-1}(\omega,\Omega)\equiv  \Phi(P_m(\overline{\Omega}))=\frac{1}{(m-1)!}\left(\frac{i}{2\pi}\right)^{m}\int_{0}^{1}dt \ \text{tr} \left(\omega\wedge \Omega_t^{m-1}\right).
\end{align}
This renders $P_m(\Omega)$ as a locally exact form, as it should by Poincar\' e's Lemma, and the associated local differential form $Q_{2m-1}(\omega,\Omega)$ is called a Chern-Simons form. Notice that, while the polynomials $P_m(\Omega)$ represent characteristic classes in the even degree cohomology of the manifold $M$, the associated Chern-Simons forms are odd degree in the cohomology of $M$. For the discussion in the present paper, the most relevant characteristic classes are the Chern class $c(E)=[\det (I +i\Omega/2\pi)]=1+ c_1(E)+ c_2(E)+...$ (the $c_{i}(E)'s$ are degree $2i$ cohomology classes and are called the Chern classes) and the Chern character $\text{ch}(E)=\left[ \text{tr}\left(e^{i\Omega/2\pi}\right)\right]= \dim E + \text{ch}_1(E)+ \text{ch}_2(E)+...$, with $\text{ch}_i(E)\equiv [P_i(\Omega)]$. In fact, the Chern class yields integer values when paired with the fundamental class of the manifold, and this is why the coefficient of the Quantum Hall response is quantized. We remark that the characteristic classes are natural with respect to pullbacks, meaning if $\phi:N\rightarrow M$ is a smooth map, we have $\text{ch}(\phi^{*}E)=\phi^{*}\text{ch}(E)$ and similarly for all characteristic classes. This property, in turn, shows that if we determine representatives of the characteristic classes for the bundle $E_0$ over $\text{Gr}_k(\mathbb{C}^n)$, then all the characteristic classes for $E=\phi^{*}E_0$ are determined by pullback by the classifying map $\phi:M\rightarrow \text{Gr}_k(\mathbb{C}^n)$.
\subsection*{Variation of the Chern-Simons forms under a gauge transformation -- relation to the Winding Classes and the Odd Chern character}
\hspace{0.5cm} Suppose we want to see how the Chern-Simons local differential form $Q_{2m-1}(\omega,\Omega)$ changes under a Gauge transformation. Namely, over the open set $U$ where the bundle is trivial, we have a local gauge transformation $g:U\rightarrow G$, where $G$ is the ``Gauge group'', giving rise rise to a local transformation $\text{pr}_1^{*}g$ over $U\times\mathbb{R}$ which, by abuse of notation, we will also denote by $g$. Then, we would like to evaluate the difference,
\begin{align}
Q_{2m-1}(\omega^{g},\Omega^{g})-Q_{2m-1}(\omega,\Omega),
\end{align}
where $\omega^g$, $\Omega^g$ are the Gauge-transformed connection and curvature forms, i.e.,
\begin{align}
\omega^{g} =g^{-1}(d+\omega) g \text{ and } \Omega^{g}=g^{-1}\Omega g.
\end{align}
For this purpose, we consider the Gauge-transformed path of connections
\begin{align}
\omega_t^{g}=g^{-1}dg +tg^{-1}\omega g=(1-t)g^{-1}dg +tg^{-1}(d+\omega) g ,
\end{align}
which is equivalent to a path between the Gauge-transformed trivial connection, $g^{-1}dg$, and the Gauge-transformed connection $\omega^g$. The homotopy formula for $Q_{2m-1}(\overline{\omega},\overline{\Omega})$ along this path then yields
\begin{align}
Q_{2m-1}(\omega^{g},\Omega^{g})-Q_{2m-1}(g^{-1}dg,0)=(d\Phi+\Phi d)Q_{2m-1}(\overline{\omega}^{g},\overline{\Omega}^{g}).
\end{align}
Because, by definition and by conjugation invariance, $dQ_{2m-1}(\overline{\omega}^{g},\overline{\Omega}^{g})=P_{2m}(\overline{\Omega}^{g})\equiv P_{2m}(\overline{\Omega})$, we have
\begin{align}
\Phi(dQ_{2m-1}(\overline{\omega},\overline{\Omega}))=\Phi(P_{2m}(\overline{\Omega}))\equiv Q_{2m-1}(\omega,\Omega),
\end{align}
where we used the definition of the Chern-Simons form again. Hence, we find,
\begin{align}
Q_{2m-1}(\omega^g,\Omega^g)-Q_{2m-1}(\omega,\Omega)=Q_{2m-1}(g^{-1}dg,0) + \text{exact form}.
\end{align}
Thus, up to exact terms, the Gauge variation of the Chern-Simons action is given by, replacing the path of connections $tg^{-1}dg$ and curvatures $(t^2-t)(g^{-1}dg)^{2}$ in the definition,
\begin{align}
Q_{2m-1}(g^{-1}dg,0)=\frac{1}{(m-1)!}\left(\frac{i}{2\pi}\right)^{m}\int_{0}^{1}dt\ (t^2-t)^{m-1} \text{tr}\left[\left(g^{-1}dg\right)^{2m-1}\right].
\end{align}
The $t$ integral can be done exactly since
\begin{align}
\int_{0}^{1} (t^2-t)^{m-1}&=(-1)^{m-1}\int_{0}^1 dt \ t^{m-1}(1-t)^{m-1}\equiv(-1)^{m-1}B(m-1,m-1)\\
&\equiv(-1)^{m-1}\frac{(m-1)!(m-1)!}{(2m-1)!},\nonumber
\end{align}
where $B(x,y)$ is the Beta function which equals the RHS for the specific integer values. Hence, we obtain,
\begin{align}
Q_{2m-1}(g^{-1}dg,0)=(-1)^{m-1}\left(\frac{i}{2\pi}\right)^{m}\frac{(m-1)!}{(2m-1)!}\text{tr}\left[\left(g^{-1}dg\right)^{2m-1}\right].
\end{align}
This brings us to the relation of the Chern characters to the winding classes of the automorphism group of a vector space. The de Rham cohomology of the automorphism group of $\mathbb{C}^k$ is the same as that of a maximal compact subgroup which can be taken to be the unitary group $\text{U}(k)$. Because the unitary group is compact, it has a bi-invariant volume form which allows us to represent cohomology classes by left and right invariant differential forms. In fact, the cohomology group of the unitary group $\text{U}(k)$, or equivalently, of $\text{GL}(\mathbb{C}^{k})$, is an exterior algebra generated by the bi-invariant forms of odd degree
\begin{align}
\omega_{2l-1}=(-1)^{l-1}\left(\frac{i}{2\pi}\right)^{l}\frac{(l-1)!}{(2l-1)!}\text{tr}\left[\left(g^{-1}dg\right)^{2l-1}\right],\ l\in \{0,...,m\},
\end{align}
namely, $H^{*}\{\text{GL}(\mathbb{C}^k)\}\cong H^{*}\{\text{U}(k)\}=\Lambda(\omega_1,...,\omega_{2k-1})$. Clearly $g^{*}\omega_{2l-1}=Q_{2l-1}(g^{-1}dg,0)$. The reason to call the classes associated to the Gauge variation of the Chern-Simons terms winding classes is then due to the following argument. The homotopy group $\pi_{2n-1}\{\text{GL}(\mathbb{C}^k)\}$ is isomorphic to $\mathbb{Z}$ provided $k\geq n$, and the isomorphism is obtained by taking a map $f:S^{2n-1}\rightarrow \text{GL}(\mathbb{C}^k)$ to the integer
\begin{align}
\int_{S^{2n-1}}f^{*}\omega_{2n-1}.
\end{align}
The polynomial associated to the total Chern character can, locally, be written as the differential of a Chern-Simons form, namely,
\begin{align}
\text{tr} \left[\exp\left(\frac{i\Omega}{2\pi}\right)\right]=\sum_{m=0}^{\infty} P_m(\Omega)=\sum_{m=1}^{\infty} dQ_{2m-1}(\omega,\Omega)\equiv dQ(\omega,\Omega),
\end{align} 
with
\begin{align}
Q(\omega,\Omega)\equiv \sum_{m=1}^{\infty}Q_{2m-1}(\omega,\Omega)=\sum_{m=0}^{\infty}Q_{2m+1}
\end{align}
By the previous arguments, we get,
\begin{align}
Q(\omega^{g},\Omega^{g})-Q(\omega,\Omega)&=Q(g^{-1}dg,0)=\sum_{m=0}^{\infty}Q_{2m+1}(g^{-1}dg,0)\\
&=\sum_{m=0}^{\infty}(-1)^{m}\left(\frac{i}{2\pi}\right)^{m+1}\frac{m!}{(2m+1)!}\text{tr}\left[\left(g^{-1}dg\right)^{2m+1}\right].\nonumber
\end{align}
Notice that the differential form on the RHS is closed and it is written in terms of the pullback by $g$ of generators of $H^{*}\{\text{GL}(\mathbb{C}^n)\}$. Thus, we have a way to associate to a map $g: M\rightarrow \text{GL}(\mathbb{C}^k)$ a cohomology class, namely $[Q(g^{-1}dg,0)]$, which only depends on the homotopy class of $g$. This cohomology class is called the odd Chern character \cite{Getzler.1993} and it is denoted by $\text{Ch}(g)$. Clearly, for maps $f:S^{2n-1}\rightarrow \text{GL}(\mathbb{C}^k)$, we have $\int_{S^{2n-1}} \text{Ch}(f)=\int_{S^{2n-1}}f^{*}\omega_{2n-1}$ thus the odd Chern character  provides the isomorphism mentioned previously $\pi_{2n-1}\{\text{GL}(\mathbb{C}^k)\}\cong\mathbb{Z}$.\\
\indent There is a natural construction which gives an explicit relation between the Chern characters and the odd Chern characters associated to the winding classes (mentioned in \cite{Bott.1978}), which can also be taken as a definition. One begins by making the observation that there is a fundamental map in the homotopy theory of unitary groups that induces an isomorphism of homotopy groups $\pi_{n}\{\text{U}(k)\}\cong \pi_{n+1}\{\text{Gr}_{k}(\mathbb{C}^k\otimes \mathbb{C}^2)\}$, for $k\geq n$. To understand this map, recall that if $X$ is a topological space, the suspension of $X$, denoted $\Sigma(X)$, is the space obtained from $X\times [0,1]$  by collapsing $X\times\{0\}$ to a point and collapsing $X\times\{1\}$ to another point. Clearly $\Sigma(S^{n})=S^{n+1}$. Also, if we have a continuous map $f:X\rightarrow Y$, there exists an associated map between suspensions $\Sigma(f):\Sigma(X)\rightarrow \Sigma(Y)$ (explicitly represented by $f\times \text{id}:X\times [0,1]\rightarrow Y\times [0,1]$, then $\Sigma(f):[(x,t)]\rightarrow [(f(x),t)]$). The suspension $\Sigma(X)$ can also be seen as two cones $C^{+}(X)$ and $C^{-}(X)$, each of them homeomorphic to $X\times [0,1]$ with $X\times \{1\}$ collapsed to a point, glued by the homeomorphic bases ($X\times \{0\}\subset C^{+}(X),C^{-}(X)$). The aforementioned map is a map $\sigma:\Sigma\left\{\text{U}(k)\right\}\rightarrow \text{Gr}_k(\mathbb{C}^{k}\otimes \mathbb{C}^2)$,
then, for each map $f:S^{n}\rightarrow \text{U}(k)$ the composition $\sigma\circ\Sigma(f):S^{n+1}\rightarrow \text{Gr}_k(\mathbb{C}^k\otimes \mathbb{C}^2)$ provides an isomorphism of homotopy groups. The map $\sigma$ is built as follows. First, in the suspension construction we take instead of $[0,1]$ the interval $[0,\infty]$. Then $\sigma$ can be obtained by the assignment
\begin{align}
\text{U}(k)\times [0,\infty]\ni(U,t)\mapsto \text{graph}(tU)=\{(w_1,w_2)\in \mathbb{C}^{k}\times \mathbb{C}^k: w_2=U\cdot w_1\}.
\end{align}
We define $e_1\equiv(1,0)$ and $e_2\equiv(0,1)$ and identify $\mathbb{C}^k\times\mathbb{C}^{k}\cong \mathbb{C}^k\otimes \mathbb{C}^2$ with $\mathbb{C}^2=\text{span}_{\mathbb{C}}\{e_1,e_2\}$. Then, clearly, we can write the following representations of $\text{graph}(tU)$
\begin{align}
\text{graph}(tU)&=\{w\otimes e_1+tU\cdot w\otimes e_2:w\in\mathbb{C}^k \}\\
&=\{(1/t)U^{-1}\cdot w\otimes e_1+w\cdot e_2:w\in\mathbb{C}^k\}.\nonumber
\end{align}
any of these explicitly rendering $\text{graph}(tU)\in\text{Gr}_k(\mathbb{C}^k\otimes \mathbb{C}^2)$. The first representation allows us to see that, when $t=0$, $\text{graph}(tU)=\mathbb{C}^{k}\otimes\text{span}_{\mathbb{C}}\{e_1\}$, independently of $U\in\text{U}(k)$, hence $\text{U}(k)\times \{0\}$ gets collapsed into a point. The second representation allows us to see that, when $t=\infty$, $\text{graph}(tU)=\mathbb{C}^{k}\otimes\text{span}_{\mathbb{C}}\{e_2\}$, independently of $U\in\text{U}(k)$, hence $\text{U}(k)\times \{\infty\}$ gets collapsed into a point. This shows that the previous assignment really does define a map from $\Sigma\left\{\text{U}(k)\right\}$ to $\text{Gr}_k(\mathbb{C}^k\otimes \mathbb{C}^2)$. Now, the map $\sigma:\Sigma\left\{\text{U}(k)\right\}\rightarrow \text{Gr}_k(\mathbb{C}^{k}\otimes \mathbb{C}^2)$ allows us to defined a vector bundle over $\Sigma\left\{\text{U}(k)\right\}$, namely $\sigma^{*}(E_0)$, where $E_0$ is the tautological $k$-plane bundle over $\text{Gr}_k(\mathbb{C}^k\otimes \mathbb{C}^2)$. Over each cone $C^{+}\{\text{U}(k)\}$ and $C^{-}\{\text{U}(k)\}$, the bundle will be trivial because these spaces are contractible. We can identify the cone $C^{+}\{\text{U}(k)\}=\{[(U,t)]\in \Sigma\{\text{U}(k)\}: t\leq 1\}$ and $C^{-}\{\text{U}(k)\}=\{[(U,t)]\in \Sigma\{\text{U}(k)\}: t\geq 1\}$. Clearly, the overlap is $\text{U}(k)\times \{1\}$. Let $f=[v_1,...,v_k]$ be the standard frame of $\mathbb{C}^{k}$, i.e., $v_i=(0,...,1,...0)$, $i\in\{1,...,k\}$. Then, we can build a local frame field over $C^{+}\{\text{U}(k)\}$, given by the local sections,
\begin{align}
v_i^{+}([(U,t)])\equiv v_i\otimes e_1+t U\cdot v_i\otimes e_2,\  i\in\{1,...,k\},\ [(U,t)]\in C^{+}\{\text{U}(k)\}.
\end{align}
Similarly, over $C^{-}\{\text{U}(k)\}$, we can use the local sections
\begin{align}
v_i^{-}([(U,t)])\equiv (1/t)U^{-1}\cdot v_i\otimes e_1+ v_i\otimes e_2,\  i\in\{1,...,k\},\ [(U,t)]\in C^{-}\{\text{U}(k)\}.
\end{align}
Clearly if $[u^{i}_j]_{1\leq i,j\leq k}$ is the matrix representing $U$ in the standard basis, we have, over $C^{+}\{\text{U}(k)\}\cap C^{-}\{\text{U}(k)\}=\text{U}(k)\times \{1\}$
\begin{align}
v_i^{+}([(U,1)])=U\cdot (U^{-1}\cdot v_i\otimes e_1+ v_i\otimes e_2)=\sum_{j=1}^{k}u_{i}^{j}v_j^{-}([(U,1)]), 
\end{align}
This means exactly that the transition map is given by the identity map in $\text{U}(k)$. The bundle can be visualized in the following figure.
\begin{figure}
\centering
\includegraphics[scale=0.5]{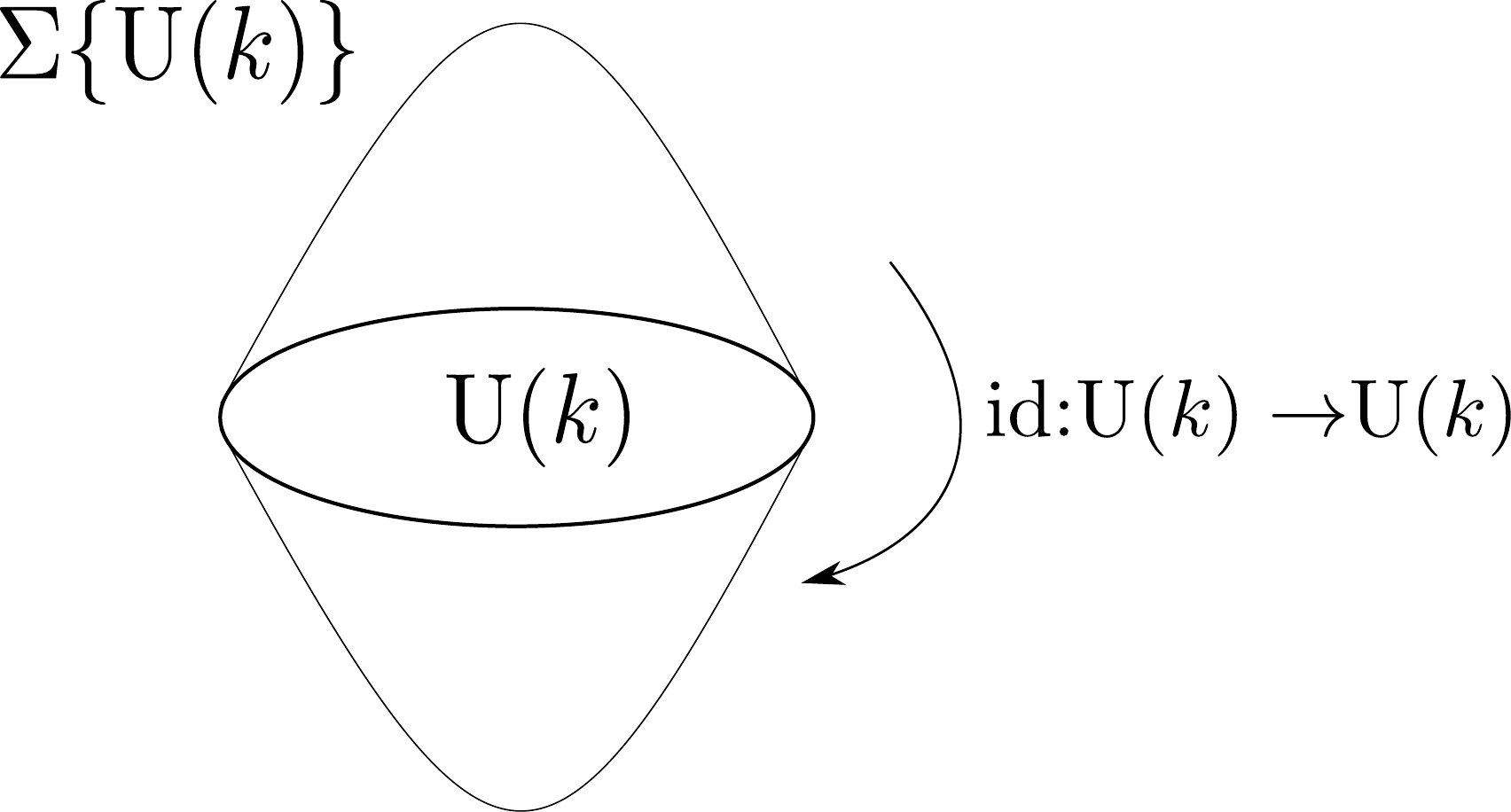}
\caption{Visualization of the pullback bundle $\sigma^{*}(E_0)\rightarrow \Sigma\{\text{U}(k)\}$: the transition map in the intersection of the cones is the identity map $\text{id}:\text{U}(k)\rightarrow\text{U}(k)$.}
\label{fig:1}
\end{figure}
 Now, introduce a connection on $\sigma^{*}(E_0)$ and denote by $\omega^{\pm}$ and $\Omega^{\pm}$ the local connections and curvatures, then, over $C^{\pm}\{\text{U}(k)\}$ , because these neighbourhoods are contractible, the differential forms representing $\text{ch}\{\sigma^{*}(E_0)\}$ are exact and given explicitly by $dQ(\omega^{\pm},\Omega^{\pm})$, but $Q(\omega^{+},\Omega^{+})-Q(\omega^{-},\Omega^{-})$ restricted to the overlap $C^{+}\{\text{U}(k)\}\cap C^{-}\{\text{U}(k)\}$ is precisely $\text{Ch}\left\{\text{id}:\text{U}(k)\rightarrow \text{U}(k)\right\}$. If $\delta: H^{\bullet}(\text{U}(k))\rightarrow H^{\bullet+1}(\Sigma\{\text{U}(k)\})$ denotes the boundary map (the suspension) in the Mayer-Vietoris sequence in cohomology, the previous statement can be compactly written as
 \begin{align}
\text{ch}\left\{\sigma^{*}(E_0)\right\}=\delta \text{Ch}\left\{\text{id}:\text{U}(k)\rightarrow\text{U}(k)\right\}.
\end{align}
In the case of maps $f:S^{2n-1}\rightarrow U(k)$ we get an important relation. If we take the suspension $\Sigma(f):S^{2n}\rightarrow \Sigma\{\text{U}(k)\}$, then we get an associated bundle $(\sigma\circ\Sigma(f))^{*}E_0\rightarrow S^{2n}$, with the transition map now on the equator $S^{2n-1}$ given by $f:S^{2n-1}\rightarrow\text{U}(k)$. Then, by the previous formula,
\begin{align}
\text{ch}\left\{(\sigma\circ\Sigma(f))^{*}E_0\right\}=\Sigma(f)^{*}\delta \text{Ch}(\text{id})=\delta \text{Ch}(f),
\end{align}
i.e., the top Chern-class is the suspension of the winding class $f^{*}\omega_{2n-1}$ and its evaluation gives the homotopy class of $f$. This result is in agreement with the fact that vector bundles over spheres of a fixed rank are in one-to-one correspondence with homotopy classes of maps from a sphere in one lower dimension to the general linear group of the given rank.\\
\indent Finally an important remark is in order. In this paper, another relation between the standard Chern characters and the odd Chern characters is yet identified. Suppose we are given a smooth map $H:X\rightarrow \text{Gr}_k(\mathbb{C}^n)$, for large $n$, where $X$ is a $2d$ dimensional manifold and where $\text{Gr}_k(\mathbb{C}^n)$ is identified as the set of unitary matrices:
\begin{align}
\text{Gr}_k(\mathbb{C}^n)\cong \{u\in \text{U}(n): u^2=1 \text{ and } u \text{ has } k \text{ negative eigenvalues}\}.
\end{align}
In particular, in our case $X=T^{2d}$ and $H$ is the Hamiltonian in momentum space. Associated with $H$, we can build a smooth map $K(H):\mathbb{R}\times X\rightarrow \text{GL}(\mathbb{C}^n)$, given explicitly by,
\begin{align}
K(H): (t, x)\mapsto i t I_{n}+H(x)
\end{align}
We use a different convention for the sign of the ``energy'' $t$ variable, in order to simplify the formulae. This map can also be understood as a composition. First we introduce the map $K_{k}:\mathbb{R}\times\text{Gr}_{k}(\mathbb{C}^n)\rightarrow \text{GL}(\mathbb{C}^n)$ given by
\begin{align}
K_{r}:(t,u)\mapsto it I_n +u.
\end{align}
Clearly, he have $K(H)\equiv K_{r}\circ(\text{id}_{\mathbb{R}}\times H)$. Then $(K(H))^{*}\omega_{2d+1}$ is a closed differential form decaying rapidly at infinity. What we have proven in the main text, is that through integration over the fibre, i.e., $\int_{\mathbb{R}}:\Omega^{\bullet}(X\times \mathbb{R})\rightarrow \Omega^{\bullet-1}(X)$, we have the relation,
\begin{align}
\int_{\mathbb{R}}\left\{K(H)\right\}^{*}\omega_{2d+1}=\text{ch}_{2d}\left\{H^{*}(E_0)\right\}=H^{*}\text{ch}_{2d}(E_0), \text{ for any } d\geq 1,
\end{align}
or,
\begin{align}
\int_{\mathbb{R}} (\text{id}_{\mathbb{R}}\times H)^{*}K_{k}^{*}\omega_{2d+1}=H^{*}\int_{\mathbb{R}}K_{k}^{*}\omega_{2d+1}= H^{*}\text{ch}_{2d}(E_0), \text{ for any } d\geq 1.
\end{align}
The above equation is true for every smooth map $H:X\rightarrow \text{Gr}_k(\mathbb{C}^n)\subset \text{U}(n)$ and for every $X$ of dimension $2d$, $d\geq 1$. This means that for $d\geq 1$, we can remove the $H$ in the formula, and simply write,
\begin{align}
\int_{\mathbb{R}} K_{k}^{*}\omega_{2d+1}= \text{ch}_{2d}(E_0), \text{ for any } d\geq 1.
\end{align}
One is then tempted to guess an equality
\begin{align}
\label{eq: Ch(K_k)-ch}
\int_{\mathbb{R}} \text{Ch}\left\{K_k: \text{Gr}_k(\mathbb{C}^n)\times\mathbb{R}\rightarrow \text{GL}(\mathbb{C}^n)\right\}=\text{ch}(E_0)
\end{align}
However, for $d=0$, we find a different result. It is natural to expect that one can introduce a ``prescription'' in the $d=0$ expression such that Eq.\eqref{eq: Ch(K_k)-ch} holds. In $d=0$, we have, explicitly,
\begin{align}
\int_{\mathbb{R}}(\text{id}_{\mathbb{R}}\times H)^{*}K_{k}^{*}\omega_{1}=\frac{i}{2\pi}\int_{\mathbb{R}}\left\{\frac{dt}{t -i}(n-k) +\frac{dt}{t+i}k \right\}.
\end{align}
We can evaluate each of the integrals in the RHS by standard complex contour integral techniques. Take for instance the first, $\int_{-\infty}^{\infty} dt/(t-i)$. If we take as a contour in the complex plane the usual semi-circle based on the real line with radius $R$ (on the upper plane for instance, containing the pole), with the radius going to infinity, we find that, applying the Residues theorem,
\begin{align}
\int_{-\infty}^{\infty}\frac{dt}{t-i}=\pi i
\end{align}
Similarly,
\begin{align}
\int_{-\infty}^{\infty}\frac{dt}{t+i}=-\pi i
\end{align}
Hence, we would find,
\begin{align}
\int_{\mathbb{R}}(\text{id}_{\mathbb{R}}\times H)^{*}K_{k}^{*}\omega_{1}=k-\frac{n}{2}
\end{align}
Since we want to obtain the previous formula, we regularize the integration over the fibre in the form
\begin{align*}
\lim_{\delta\rightarrow 0^{+}}\frac{i}{2\pi}\int_\mathbb{R} \left\{\frac{dt}{t-i}(n-k)+\frac{dt}{t+i}k\right\}e^{i t \delta}, 
\end{align*} 
in this case we see that only the first integral survives (taking the contour to be the semicircle on the lower half plane based on the real line, there is now a damping term), and we get the interesting result: 
\begin{align}
\int_{\mathbb{R}} \text{Ch}\left\{K_k: \text{Gr}_k(\mathbb{C}^n)\times\mathbb{R}\rightarrow \text{GL}(\mathbb{C}^n)\right\}=-(n-\text{ch}(E_0))=-\text{ch}(E_0^{\perp}),
\end{align}
where $E_0^{\perp}$ is the orthogonal complement bundle to the tautological $k$-plane bundle $E_0\subset \text{Gr}_k(\mathbb{C}^n)\times\mathbb{C}^n$ (of course, in the LHS, $\int_{\mathbb{R}}$ should be understood with the regularization discussed). This is still different from what we guessed. If we regularize it as
\begin{align*}
\lim_{\delta\rightarrow 0^{+}}\frac{i}{2\pi}\int_\mathbb{R} \left\{\frac{dt}{t-i}(n-k)+\frac{dt}{t+i}k\right\}e^{-i t \delta}, 
\end{align*}
we get the desired result 
\begin{align*}
\int_{\mathbb{R}} \text{Ch}\left\{K_k: \text{Gr}_k(\mathbb{C}^n)\times\mathbb{R}\rightarrow \text{GL}(\mathbb{C}^n)\right\}=\text{ch}(E_0).
\end{align*}
Notice that this regularizations do not affect the higher dimensional ($d\geq 1$) cases, so it is a matter of convenience. We could also just adopt the symmetric formula,
\begin{align*}
\int_{\mathbb{R}} \text{Ch}\left\{K_k: \text{Gr}_k(\mathbb{C}^n)\times\mathbb{R}\rightarrow \text{GL}(\mathbb{C}^n)\right\}=\text{ch}(E_0)-\frac{n}{2}.
\end{align*} 
We argue that the topological response should not depend on $n$, since one can add an arbitrary number of trivial flat bands and the response should be the same. Therefore, we adopt the second regularization. Ultimately, the reason why this choice is physical can probably be traced back to the time ordering of the fields, but we have decided not to pursue this direction.
\subsection*{Chern-Simons forms for Abelian Gauge fields}
\hspace{0.5cm} For the particular case of an Abelian Gauge field, we set $\omega\equiv A$ and $\Omega\equiv F$ and solving the integral defining $Q_{2m-1}$, we get the simplified expression appearing in the main text:
\begin{align}
Q_{2m-1}(A,F)&=\frac{i^m}{(m-1)!(2\pi)^m}\int_{0}^{1}dt\ t^{m-1}A\wedge F^{m-1}\\
&=\frac{i^m}{m!(2\pi)^m}A\wedge F^{m-1}.\nonumber
\end{align}
\section*{Appendix B: Anomalies and descent relations}
\hspace{0.5cm} When we have a symmetry at the level of the classical action of a theory, it may happen that this symmetry does not hold at the quantum level. Namely, the measure in the path integral may have an anomalous variation and this leads to an anomalous current flow. One such example is the case of a gauge symmetry. Suppose we have a theory for fermions in some representation $r$ of a Gauge group $G$. They by integrating over the fermions we obtain an effective action for the Gauge field $W(A)$. An anomaly occurs if when we perform an infinitesimal Gauge transformation the divergence of the current is non-vanishing. Namely, if the Gauge transformation is parameterized by an infinitesimal group element $g=1+v$ then
\begin{align}
A\rightarrow A^{g}=g^{-1}(d+A)g=A+\delta_v A\equiv A+Dv,
\end{align}
 with $Dv=dv +[A,v]$ being the Gauge covariant derivative, and we have,
\begin{align}
\delta_v W(A)=G(v,A),
\end{align}
for some non-vanishing functional $G(v,A)$, so the effective action is not invariant. The Gauge anomalies satisfy consistency conditions, called the Wess-Zumino consistency conditions \cite{Nakahara.2003}, namely,
\begin{align}
\delta_{v_1}\delta_{v_2}W(A)-\delta_{v_2}\delta_{v_1}W(A)-\delta_{[v_1,v_2]}W(A)=0,
\end{align}
for any two infinitesimal Gauge parameters $v_1,v_2$. This consistency conditions is analogous to the relation $d^2=0$ for the exterior derivative $d$. The Becci-Rouet-Stora (BRS) operator $s$ is an exterior derivative-like operator on the space of connections which commutes with $d$ ($ds+s d\equiv 0$). The previous equation reads $s^2 W(A)\equiv 0$. It is useful to introduce the Fadeev-Popov ghost $v\equiv g^{-1}s g$, where $g$ is the Gauge transformation parameterizing the  Gauge field $A^{g}=g^{-1}(d+A)g$. Then, we can write,
\begin{align}
s W(A)=G(v,A).
\end{align} 
By starting from the Gauge invariance of the Chern character associated to the Chiral or Abelian anomaly and the index of a Dirac operator in flat space-times, it is possible to find the so-called descent relations for the coefficients of the expansion of the Chern-Simons form
\begin{align}
Q_{2m+1}(A+v,F)=Q_{2m+1}^{0}(A,F)+Q_{2m}^{1}(v,A,F)+...+Q^{2m+1}_{0}(v,A,F),
\end{align}
namely,
\begin{align}
s Q^{0}_{2m+1}(A,F)+dQ^{1}_{2m}(v,A,F)=0,\\
s Q^{1}_{2m}(v,A,F)+dQ^{2}_{2m-1}(v,A,F)=0, \nonumber\\
\vdots \nonumber
\\
s Q^{2m}_{1}(v,A,F)+dQ^{2m+1}_{0}(v,A,F)=0, \nonumber\\
sQ^{2m+1}_0(v,A,F)=0.\nonumber
\end{align}
Now let $M^{2m}$ be a closed $2m$ dimensional manifold over which we have a gauge field $A$. Taking $G(v,A)\equiv \int_{M^{2m}} Q_{2m}^{1}(v,A,F)$, we have,
\begin{align}
s G(v,A)&=s\int_{M^{2m}} Q_{2m}^{1}(v,A,F)=\int_{M^{2m}}s Q_{2m}^{1}(v_2,A,F)\\
&=-\int_{M^{2m}}d Q_ {2m-1}^{2}(v,A,F)=0, \nonumber
\end{align}
since $\partial M^{2m}=\emptyset$. Hence $G(v,A)$ is consistent with an anomaly for an effective action $W(A)$.
\bibliographystyle{aipnum4-1}
\bibliography{bib}
\end{document}